\documentclass[symmetry,review,accept,pdftex,moreauthors]{Definitions/mdpi} 
\usepackage{wasysym}
\usepackage{physics}
\usepackage{aas_macros}  %to resolve biblio macro errors
\usepackage{longtable}  %for tables of more than one page
\usepackage{booktabs}   %for \midrule and \toprule in tables

\newcommand{\mdy}[1]{{#1}}%{\textbf{#1}

\firstpage{1} 
\pubvolume{1}
\issuenum{1}
\articlenumber{0}
\pubyear{2024}
\copyrightyear{2024}
\datereceived{25 Nov 2024 } 
\daterevised{21 Dec 2024 } % Comment out if no revised date
\dateaccepted{25 Dec 2024 } 
\datepublished{ } 
\hreflink{https://doi.org/} % If needed use \linebreak
\usepackage{subcaption}

\Title{Circumstellar and circumbinary discs in multiple stellar systems}

\TitleCitation{Discs in multiple stellar systems}

\Author{Nicol\'as Cuello $^{1,\dagger}$\orcidA{}, Antoine Alaguero $^{1,\dagger}$\orcidB{} and Pedro P. Poblete $^{1}$}

\AuthorNames{Nicol\'as Cuello, Antoine Alaguero and Pedro Poblete}

\AuthorCitation{Cuello, N.; Alaguero, A.; Poblete, P.P.}
\address[1]{%
$^{1}$ \quad Univ. Grenoble Alpes, CNRS, IPAG, 38000 Grenoble, France}

\corres{Correspondence: nicolas.cuello@univ-grenoble-alpes.fr, antoine.alaguero@univ-grenoble-alpes.fr}

\firstnote{These authors contributed equally to this work.}  % Current address should not be the same as any items in the Affiliation section.
\abstract{The interplay between stellar multiplicity and protoplanetary discs represents a cornerstone of modern astrophysics, offering key insights into the processes of planet formation. Protoplanetary discs act as cradles for planetary systems, yet their evolution and capacity to form planets are profoundly affected by gravitational forces within multiple stellar systems. This review synthesises recent advancements in observational and theoretical studies to explore the rich diversity of circumstellar and circumbinary discs within multiple stellar systems. We examine how stellar companions shape disc morphology through truncation, spirals, and misalignment. We also outline how dust dynamics and planetesimal formation are impacted by stellar multiplicity. On top of this, observations at high angular resolution reveal detailed disc structures, while simulations offer key insights into their evolution. Last, we consider the implications of stellar multiplicity for planetary system architectures, emphasising the diversity of planetary outcomes in such environments. Looking ahead, coordinated efforts combining high-resolution observations with advanced numerical models will be critical for unravelling the role of multiple stellar systems in shaping planetary formation and evolution.}


\begin{document}

%%%%%%%%%%%%%%%%%%%%%%%%%%%%%%%%%%%%%%%%%%
\section{Introduction}

The formation of stars and planets is intimately linked through the physical processes that unfold within protoplanetary discs, which are rotating structures of gas and dust that surround young stellar objects \citep{Manara+2023}. These discs are the cradles of planetary systems where microscopic dust grains are transformed into planetesimals, which eventually evolve into rocky or gaseous planets --- depending on their surrounding conditions and migration \citep{Armitage2020}. The dynamics, structure, and lifetime of these discs are key in determining the diversity of planetary architectures observed in the galaxy. However, a definitive link between protoplanetary discs and the resulting exoplanets has yet to be firmly demonstrated  \citep{Drazkowska+2023}. This topic constitutes an active topic of research in modern-day astrophysics. 

In addition, it is now well established that multiple stellar systems are ubiquitous, particularly among young stars \citep{Offner+2023}. Observational surveys reveal that \mdy{around $65\%$} of stars are born not in isolation but as members of binary or higher-order multiple systems \citep{DucheneKraus2013, Reipurth+2014}. In fact, the prevalence of stellar multiplicity raises fundamental questions about its influence on protoplanetary discs. How do interactions between stellar companions shape the distribution and evolution of gas and dust? To what extent do these interactions enhance or hinder planet formation? Are planetary systems in multiples different from the ones around single stars? All in all, multiple stellar systems represent astrophysical laboratories of great scientific value if one wishes to decipher the process of planet formation. 

\begin{figure}[H]
\begin{center}
\includegraphics[width=\textwidth]{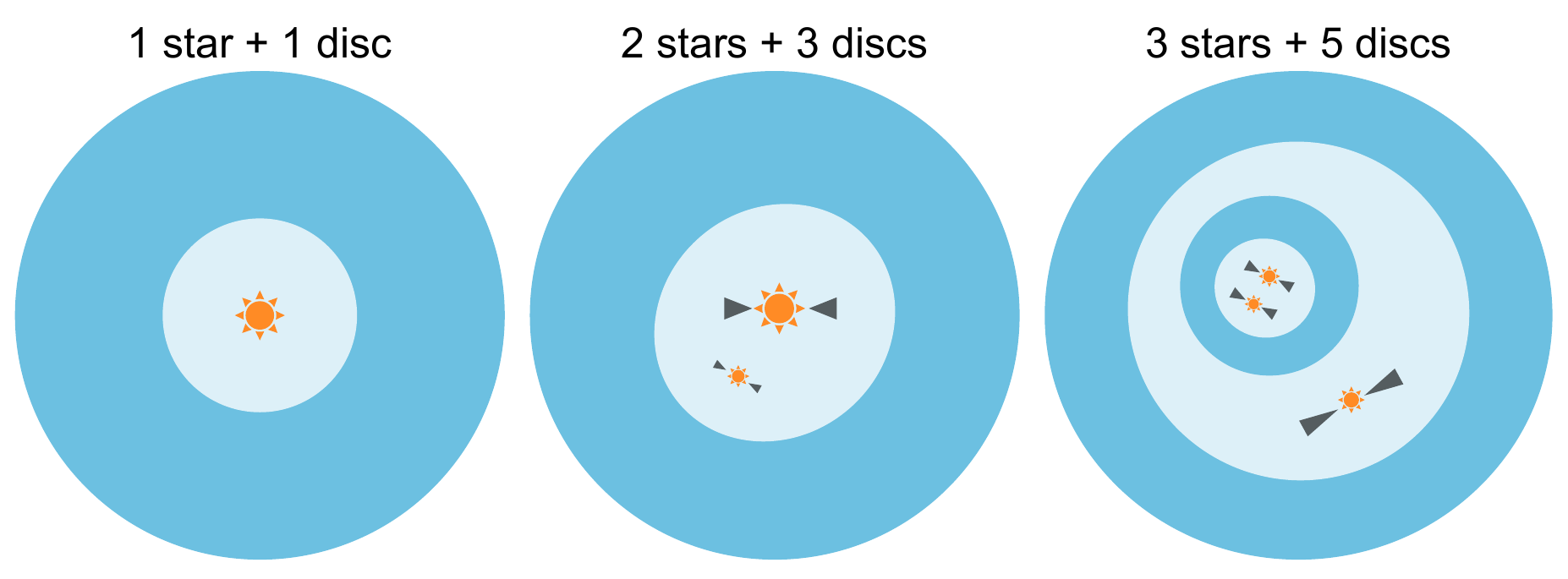}
\end{center}
\caption{Possible types of protoplanetary discs in systems with one star (left), two stars (centre), and three stars (right): circumstellar, circumbinary, or circumtriple. For simplicity, we show hierarchical stellar configurations only. For clarity, the circumstellar discs are shown in edge-on configurations.}
\label{fig:disc-types}
\end{figure}

When considering discs in multiple stellar systems, it is essential to distinguish between circumstellar discs (surrounding individual stars in a system) and circum-multiple discs (encircling two or more stars at the centre). For instance, Figure~\ref{fig:disc-types} shows the maximum number of discs within hierarchical multiple systems\footnote{A hierarchical multiple system refers to a stellar configuration in which the system can be divided into nested subsystems, each gravitationally bound and dynamically stable over long timescales. For instance, a triple system can consist of a close binary orbited by a third star at a much larger distance. The stability of such systems arises because the gravitational interactions between stars within each subsystem dominate over interactions between different subsystems.}: 1 for singles, 3 for binaries, and up to 5 for triples. These configurations differ not only in their geometrical setup but also in their formation histories, dynamical behaviour, and long-term evolution \citep{Bate2018, Tokovinin2021, Offner+2023}. Circumstellar discs may experience truncation and gravitational perturbations due to the presence of an external stellar companion \citep{Paczynski1977, MirandaLai2015}, while circumbinary and circumtriple discs are shaped by the motion of the stars within their inner cavity \citep{ArtymowiczLubow1994, Ragusa+2020}. If we consider multiple stellar systems in hierarchical arrangements, to first order, triple and quadruple systems can often be approximated as binary systems. Based on this observation, here, we will primarily address circumstellar and circumbinary discs --- mentioning when relevant the detailed orbital arrangement of a given system. 

This review aims to explore the rich diversity of protoplanetary discs in multiple stellar systems, focusing on their distinct formation pathways, dynamical interactions, and implications for planet formation. By synthesising recent observations and theoretical advances, we seek to highlight the critical role that stellar multiplicity plays in shaping the architecture of planetary systems. In Section~\ref{sec:orbitsdiscs} we present the orbital elements and typical parameters used to describe and model stars and discs in multiples. Then, in Sections~\ref{sec:CSDs} and~\ref{sec:CBDs}, we give an overview of the processes governing the dynamics in discs of multiple stellar systems and summarise recent observations of circumstellar and circumbinary discs in multiples (respectively). Finally, in Section~\ref{sec:discussion}, we discuss these observations in the context of planet formation theories, and we list our conclusions in Section~\ref{sec:conclusions}.

%%%%%%%%%%%%%%%%%%%%%%%%%%%%%%%%%%%%%%%%%%
\section{Orbital parameters and typical disc parameters}
\label{sec:orbitsdiscs}

To describe stellar orbits alongside the dynamics and structure of discs in multiple stellar systems, we first introduce key concepts in celestial mechanics (Sect.~\ref{sec:orbits}) and protoplanetary disc models (Sect.~\ref{sec:discs}). This will provide a useful framework for analysing individual disc observations and understanding the underlying physical mechanisms.

\subsection{Orbital characterisation}
\label{sec:orbits}

Traditionally, the orbital parameters used to describe the motion of a two-body system bound by gravity are defined relative to the centre of mass. These parameters are the semi-major axis ($a$), the eccentricity ($e$), the inclination ($i$), the longitude of the ascending node ($\Omega$), the argument of periapsis ($\omega$), and the true anomaly ($\nu$). All these parameters are shown in Fig.~\ref{fig:orbital-elements} and we provide a detailed explanation in Appendix~\ref{sec:app_orb}. For a specific system, the values of these parameters determine the discs' sizes and the type of alignment or oscillations that circumstellar and circumbinary discs can undergo. For instance, the value of the binary semi-major axis $a$ puts strong constraints on circumstellar disc sizes and on circumbinary disc inner cavities. The process of polar alignment of a circumbinary can only occur if the inner binary orbital eccentricity verifies $e>0$ \citep{FaragoLaskar2010, Aly+2015, MartinLubow2017}. Also, the closer $\Omega$ gets to $90^\circ$, the more likely the process of polar alignment becomes \citep{FaragoLaskar2010, CuelloGiuppone2019}. Last, the binary inclination $i$ critically determines the precession rate and oscillation evolution of misaligned discs \citep{Larwood+1996}.

\begin{figure}[H]
\begin{center}
\includegraphics[width=0.89\textwidth]{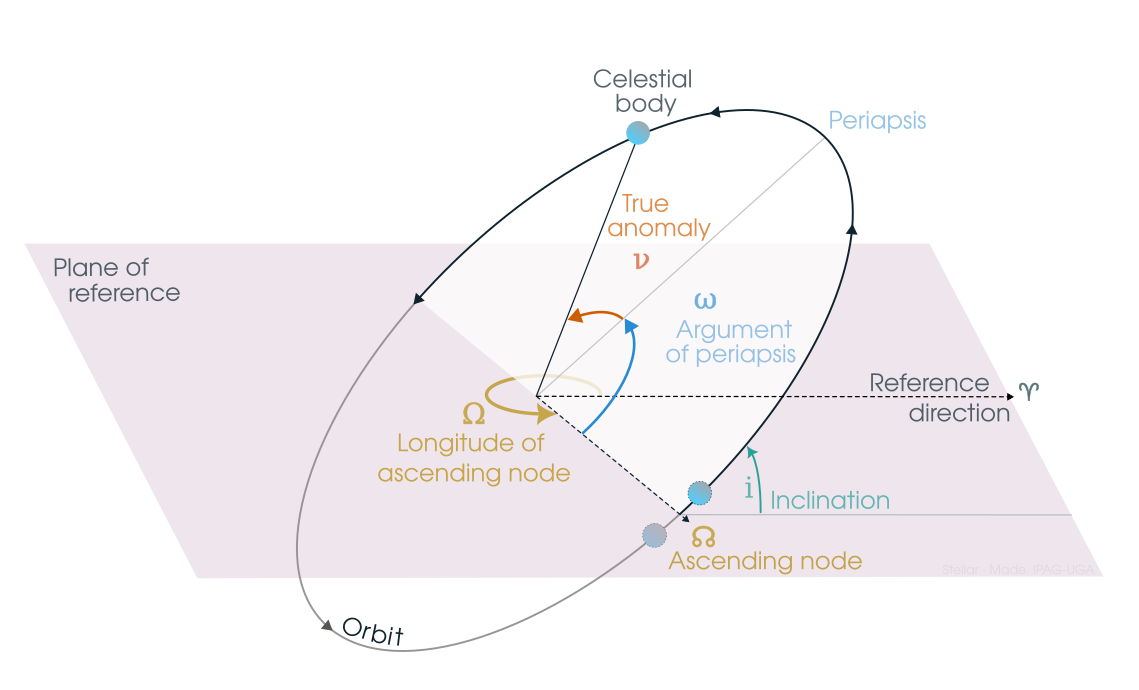}
\end{center}
\caption{Schematic illustration of the orbital parameters for an elliptical orbit: the inclination $i$, the argument of periapsis $\omega$, the longitude of the ascending node $\Omega$, and the true anomaly $\nu$. Provided through the courtesy of Mario Sucerquia.}
\label{fig:orbital-elements}
\end{figure}

The orbital parameters of multiple stellar systems can be precisely constrained through high-precision astrometric observations conducted over extended periods with instruments such as VLT/GRAVITY \citep{Gravity+2017}, VLT/SPHERE \cite{Beuzit+2019} and Karl G. Jansky Very Large Array (VLA). This detailed characterisation is critical for linking stellar motions to the structures observed in protoplanetary discs. This integrated approach not only bridges observational data with theoretical models but also enhances our understanding of how stellar orbits influence disc behaviour and, ultimately, planet formation. For instance, the values of the binary mass ratio $q$ and ($a$, $e$, $i$) strongly impact the morphology and evolution of the inner cavity of circumbinary discs \citep{Ragusa+2017, Thun+2017, Poblete+2019, Hirsh+2020, Ragusa+2021}. Relative misalignments between stellar and disc orbital planes can also indicate the oscillation regime of a given disc in a multiple stellar systems \citep{MartinLubow2017, Smallwood+2021, Ceppi+2023, Toci+2024}. Last, disc substructure in the form of spirals and over-densities can be directly related to different sets of binary orbital parameters. This last connection is better established through detailed hydrodynamical simulations for several hundreds or thousands binary orbital periods \citep{Price+2018, Thun+2017, Gonzalez+2020, LaiMunoz2023, Calcino+2024, Penzlin+2024, Ragusa+2024}. For instance, recent works on the multiple stellar system GG~Tau \cite{Duchene+2024, Toci+2024} present disc dynamics simulations done with great accuracy using updated astronomical constraints  --- shedding light on the complex ongoing interactions. In the following section, we outline the fundamental equations governing the evolution of protoplanetary discs.

\subsection{Protoplanetary discs characterisation}
\label{sec:discs}

Protoplanetary discs are rotating structures made of gas and solid material that surround young stars, with gas being the dominant component (exceeding the solid material by at least an order of magnitude, \cite{Birnstiel+2010}). This gas-dominated nature allows protoplanetary discs to be treated as fluids, requiring the use of hydrodynamical equations for accurate modelling. In this context, the Navier-Stokes equations are fundamental to describe the fluid motion around young stars --- though they remain unsolved in their general form. Therefore, we must rely on simplifying assumptions and approximations to write the governing equations of protoplanetary discs. Assuming that the fluid is incompressible and follows an adiabatic equation of state, these equations can be written as follows:
\begin{eqnarray}
    \textit{Mass conservation:} & & 
    \frac{\partial \rho}{\partial t} + \rho\left( \nabla\cdot {\bf v} \right) = 0, \\
    \textit{Momentum conservation:} & & 
    \dv{\bf v}{t} = -\frac{\nabla P}{\rho} + \Pi_{\rm shock} + \boldsymbol{a}_{\rm grav}, \\
    \textit{Energy conservation:} & & 
    \dv{u}{t} = \frac{P}{\rho^2}\dv{\rho}{t} + \Lambda_{\rm shock} - \frac{\Lambda_{\rm cool}}{\rho}, \\
    \textit{Equation of state:} & & 
    P = \left(\gamma - 1\right)\rho u.
\end{eqnarray}
Here, $\rho$ denotes the gas density, ${\bf v}$ is the local fluid velocity vector, $P$ represents the gas pressure, $u$ is the gas internal energy, and $\gamma$ is the adiabatic index. The term $\boldsymbol{a}_{\rm grav}$ accounts for external gravitational acceleration due to stars, planets, or the disc’s self-gravity. Additionally, $\Pi_{\rm shock}$ includes an effective shock-viscous contribution designed to capture discontinuities at shocks \citep{Price2012, Armitage2020}. More precisely, it is a vector arising from the divergence of a shock-viscous stress tensor, added in many numerical methods to suppress unphysical oscillations near steep gradients \citep{Price2012, Teyssier2015}. Its dissipative counterpart in the energy equation is $\Lambda_{\rm shock}$, which represents the conversion of kinetic energy into heat wherever shocks are present. Formally, $\Lambda_{\rm shock}$ is derived from the same artificial-viscosity or effective-viscosity approach used to construct $\Pi_{\rm shock}$, ensuring that momentum and energy are treated self-consistently in shock regions. Meanwhile, $\Lambda_{\rm cool}$ encompasses radiative and other cooling processes. Together, these parameters encapsulate the fundamental physical processes governing protoplanetary discs. This framework can be easily expanded to include additional terms when other physical processes are relevant (e.g., ionisation from the interstellar medium, electromagnetic effects, gas chemistry; \citep{Lesur+2023}).

With these four fundamental equations in mind, power-laws are often used to model the radial surface density and temperature profiles: $\Sigma(r) = \Sigma_0 \left(r/r_0\right)^{-p}$ and  $T(r) = T_0 \left(r/r_0\right)^{-q}$, where $r_0$ is a reference radius, and $\Sigma_0$ and $T_0$ are constants. In addition, the gas vertical density structure typically follows: $\rho(z) = \rho_0\ {\rm exp}\left( - z^2 / (2h^2) \right)$, where $z$ is the vertical cylindrical coordinate, $h$ is the scale height, $q$ is the power-law index, and $\rho_0$ is a constant. This Gaussian profile arises from the assumption of hydrostatic equilibrium in the vertical direction, where the vertical pressure gradient is balanced by the gravitational force. Mathematically, this balance leads to an exponential decline of the density with height, with $h$ determined by the thermal properties of the gas and the mass of the central star \citep{Armitage2020}. Under such assumptions, all these parameters are linked to the total mass of the disc, which is challenging to estimate observationally. the typical disc-to-star mass ratios range from 1\% to 20\% \citep{Miotello+2023, Manara+2023}, though these values may vary for discs in multiple stellar systems. A caveat of the vertical isothermal assumption is that it inherently neglects the vertical thermal stratification expected in discs, which plays a crucial role in shaping their dynamical, dust, and chemical processes \citep[e.g.][]{Villenave+2020, Paneque-Carreno+2023, Pfeil+2024}.

Gas is the most abundant component of the disc, but dust (though less prevalent) plays a pivotal role in regulating opacity, forming solid cores, and influencing chemical abundances. The exact gas-to-dust ratio in a protoplanetary disc is not precisely known but is often assumed to resemble that of the interstellar medium, with a typical dust-to-gas ratio of 0.01 \citep{Bohlin+1978}. The solid material mainly consists of silicate dust grains, which vary widely in size and shape \cite[e.g.][]{TelescoKnacke1991}. Size is particularly significant when modelling gas-dust interactions, as the grain size distribution is assumed to follow a power law with an index $m$ \mdy{defined by equation \ref{eq:grainindex} below}: 
\begin{equation}
\label{eq:grainindex}
   \dv{n}{s} = s_0 s^{-m},
\end{equation}
where $n$ is the number of particles with radius $s$, and $s_0$ is a normalisation constant. The standard value for $m$ measured in the interstellar medium is 3.5, indicating that smaller particles are more abundant than larger ones \citep{Mathis+1977}. Grain sizes range from 0.01 $\mu$m to centimetre scales, while particles exceeding metre sizes are often categorised as solid cores or embryos. The disc’s optical properties are dominated by small grains, whereas the majority of the mass resides in larger grains \citep{Miotello+2023}.

In protoplanetary discs, dust particles interact with the surrounding gas exchanging angular momentum and typically evolving under the Epstein regime, in which the mean free path of gas molecules exceeds the particle size \citep{Epstein1924}. From the physical point of view, a dust grain in motion within the disc experiences an aerodynamic drag force proportional to the velocity difference between the grain and the gas. With the gas orbiting at a sub-Keplerian frequency due to pressure forces, the dust grains orbiting at a Keplerian frequency start to drift toward the star and to settle in the mid-plane of the disc \citep{Weidenschilling1977, Laibe+2012}. The degree of coupling between dust and gas, characterized by the Stokes number ${\rm St}$, governs dust evolution and distribution within the disc. Particles strongly coupled to the gas (${\rm St} \ll 1$) follow the gas flow, while decoupled particles (${\rm St} \gg 1$) move independently in Keplerian orbits. Marginally coupled particles (${\rm St} \sim 1$) experience the maximum radial drift, a key factor in determining dust concentration and growth. Instruments like SPHERE\footnote{The Spectro-Polarimetric High-Contrast Exoplanet Research (SPHERE) instrument, mounted on the Very Large Telescope (VLT) in Chile, is optimized for high-resolution imaging and polarimetric observations, particularly in scattered light, to study circumstellar environments and exoplanets.} can trace the stellar light scattered by small grains tightly coupled to the gas \citep{Benisty+2023}, while instrument like ALMA\footnote{The Atacama Large Millimeter/submillimeter Array (ALMA) is an interferometric radio telescope located in Chile, designed for high-resolution observations of dust, gas, and other structures in protoplanetary discs and the interstellar medium.} probe the thermal emission of grains with sizes close to the millimetre \citep{Andrews2020}.

%%%%%%%%%%%%%%%%%%%%%%%%%%%%%%%%%%%%%%%%%%
\section{Circumstellar discs (CSDs) in multiple stellar systems}
\label{sec:CSDs}

Circumstellar discs in multiple stellar systems are heavily influenced by gravitational perturbations from nearby stellar companions, which profoundly impact their structure, dynamics, and evolution. In the following, we briefly outline the key mechanisms at play and present a limited but representative selection of circumstellar discs in such systems.

\subsection{Tidal truncation of CSDs}
\label{subsec:CSD_truncation}

In binary and multiple stellar systems, circumstellar discs experience tidal truncation due to the gravitational influence of nearby stellar companions. This process occurs as the torque exerted by the outer companion limits the radial extent of the disc, effectively \textit{truncating} its size \citep{Paczynski1977, MirandaLai2015}. The truncation radius depends on the mass ratio and separation of the stars, as well as the orbital configuration of the system. As a result, circumstellar discs in these systems are typically smaller than those around isolated stars \citep[e.g.][]{Manara+2019, Akeson+2019, Zurlo+2020, Zurlo+2023}. Beyond reducing the size of the disc, tidal truncation also steepens the radial surface density profile of the gas \citep{Rosotti+2018, Cuello+2019, Rota+2022}. This change in density has significant implications for the evolution of solids within the disc. A steeper surface density profile increases the pressure gradient, which in turn accelerates the radial drift of dust and larger particles toward the central star \citep{Rosotti+2018, Zagaria+2021, Zagaria+2023}. This enhanced drift can deplete the disc of solids more quickly, potentially stifling planet formation processes.

As a consequence, circumstellar discs in multiple stellar systems are thought to have shorter lifespans compared to their counterparts around single stars \citep{Cieza+2009, Alexander2012, Ronco+2021}. However, this limitation may be mitigated if the discs are resupplied with material from external reservoirs. Such reservoirs could include circumbinary discs, infalling envelopes, or streamers that channel material inward. These mechanisms may play a crucial role in sustaining circumstellar discs in the presence of tidal truncation, thereby extending their lifetimes and maintaining conditions suitable for planet formation (see Sects.~\ref{sec:CBDs} and \ref{sec:discussion}).

\subsection{Structure formation in CSDs}
\label{subsec:CSD_structures}

Gravitational interactions between a disc and a nearby stellar companion can lead to the emergence of striking structures within the disc, with spiral density waves being one of the most prominent examples \citep{Offner+2023, Cuello+2023}. These waves are launched at Lindblad resonances \citep{Lindblad1941}, where the companion’s gravitational influence perturbs the disc material, generating density waves that propagate through the gas and dust \citep{Rafikov2002}. The morphology of the spirals depends on a combination of parameters, including the companion’s mass and orbital separation, as well as the disc’s properties, such as viscosity, cooling efficiency, and thickness \citep{Forgan+2018, Muley+2024}. High-mass companions or close-in perturbers tend to produce more prominent and tightly wound spirals, while discs with higher cooling efficiency exhibit sharper and more defined features. These spirals are observable across different tracers, with scattered light revealing the fine structures in the disc’s upper layers \citep{Benisty+2023}, molecular line emission tracing gas kinematics \citep{Pinte+2023}, and dust thermal emission highlighting the denser mid-plane regions (e.g., HD 100453, \cite{Gonzalez+2020}; AS 205, \cite{Kurtovic+2018}; \cite{Andrews2020}).

The interaction between a companion and a disc not only generates spirals but can also drive material outward, where it may circularise and form transient circumstellar discs (CSDs) around the companion. This process is more likely to occur in unbound encounters, also known as flybys, where the perturber does not remain gravitationally bound to the system \citep{Borchert+2022, Smallwood+2024a}. In such cases, the companion’s interaction can eject substantial material from the disc, some of which may accumulate around the perturber. In contrast, for bound companions, the spirals are periodically excited with each orbit, although their amplitude diminishes over time as the disc is truncated and its mass redistributed \citep[e.g.,][]{Menard+2020}. Consequently, the formation of second-generation circumstellar discs around the perturber is unlikely in bound systems.

\subsection{CSD alignment}
\label{subsec:CSD_alignment}

The alignment of circumstellar discs (CSDs) in binary and multiple stellar systems is influenced by the gravitational interactions between the stars and the disc. When the orbital plane of the companion star is inclined relative to the plane of the disc, the disc experiences torques that can induce oscillations in its inclination. These oscillations occur around an equilibrium plane determined by the total angular momentum of the system, which combines the angular momenta of the stars and the disc(s). The amplitude and frequency of these oscillations strongly depend on the system’s orbital parameters: the companion’s mass, separation, inclination, and orbital eccentricity; as well as the disc mass, radial extent, viscosity, and thickness \citep{Bate+2000, LubowOgilvie2000}. 

In many configurations, the disc may begin to precess around the angular momentum vector of the entire stellar system. This precession can occur either differentially, where individual disc annuli precess at different rates, or as a solid body, where the entire disc precesses uniformly. The mode of precession depends on the balance between wave-like communication (dominated by sound speed and pressure forces) and viscous timescales within the disc \citep{PapaloizouTerquem1995}. For thin, inviscid discs, differential precession is common, with the precession rate varying as $a^{3/2}$, where $a$ is the radial distance from the central star. In contrast, thick or highly viscous discs are more likely to exhibit solid-body precession.

In systems with strongly inclined or eccentric orbits, the interaction can drive more complex dynamics. For instance, the disc may undergo Kozai-Lidov-like oscillations, where the disc inclination and eccentricity oscillate periodically due to the influence of the companion’s gravitational potential \citep{Martin+2014}. These oscillations can lead to significant warping or disc breaking if the companion’s torque overcomes the internal forces maintaining the disc’s structure \citep{Facchini+2013}. Over time, the dissipation of energy through viscosity may lead to alignment or misalignment between the disc and the orbital plane of the companion star, depending on the system’s initial configuration \citep{LubowOgilvie2000}. Viscous forces within the disc act to damp oscillations, promoting alignment with the total angular momentum vector of the system. However, in some cases, the initial conditions and ongoing perturbations can result in persistent misalignment, which is observed in several multiple stellar systems \citep{Bate2018, Czekala+2019, Ceppi+2024}.

\subsection{CSDs of interest}
\label{subsec:CSD_systems}

\begin{figure}[H]
\centering
\begin{center}
\includegraphics[width=0.7\textwidth]{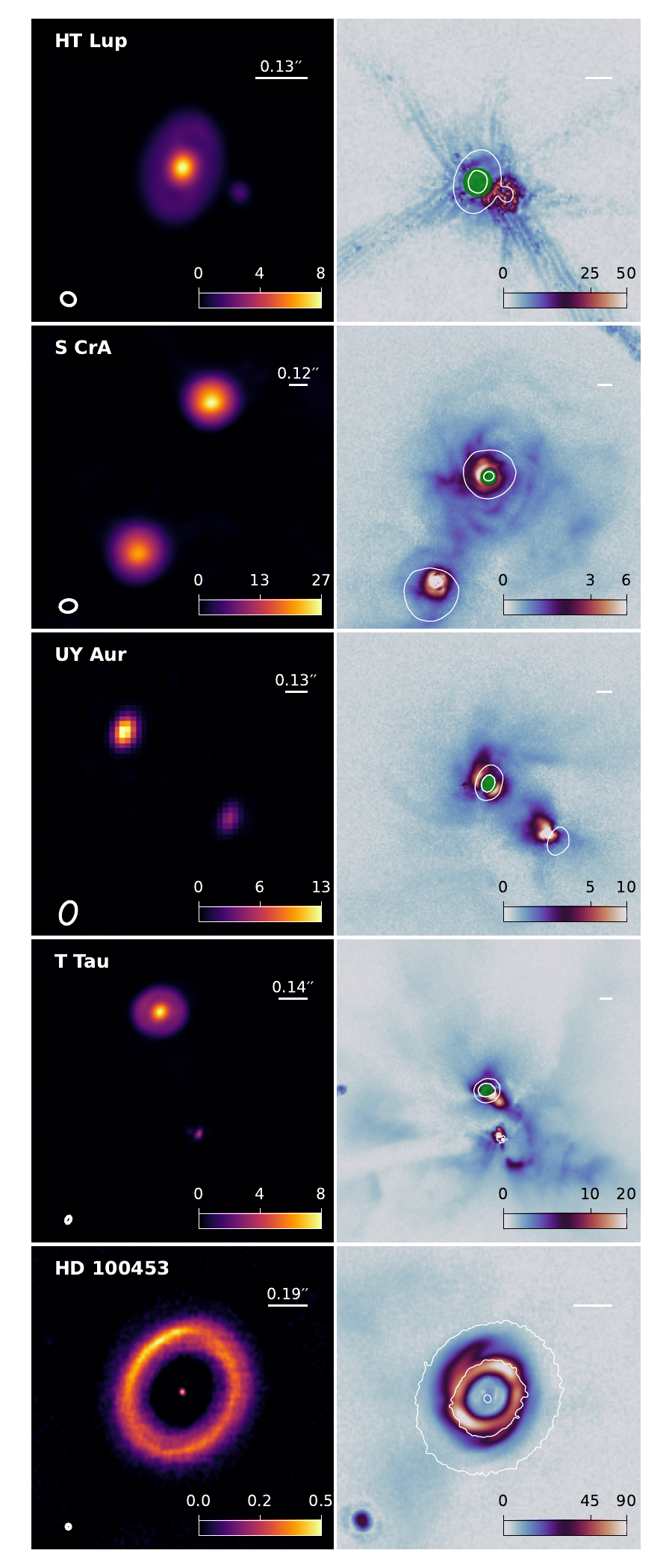}
\end{center}
\caption{
Gallery of observed circumstellar discs (CSDs) with external companions. Left column: ALMA 1.3 mm intensity maps (mJy/beam) with the synthesized beam and a 20 au scale bar (angular size noted). Right column: SPHERE polarized intensity data (\% of peak value) processed with {\sc IRDAP}, showing a 20 au scale bar and a green circle for the coronagraph position, if used. SPHERE data are at 1.65 µm, except 2.15 µm for HD 100453. White contours show ALMA 1.3 mm emission at [5, 50]$\sigma$, centred on the coronagraph or primary star.}
\label{fig:CSDs}
\end{figure}

In the following, we describe ALMA and SPHERE observations of multiple stellar systems harbouring circumstellar discs in an attempt to illustrate the interactions between a disc and an external companion. We list these systems in Table~\ref{tab:systems} and show a gallery of recent observations in Figure \ref{fig:CSDs}. The SPHERE polarimetric data were downloaded from the ESO archive\footnote{\url{http://archive.eso.org/eso/eso_archive_main.html}} and then processed using the {\sc IRDAP} open-access pipeline\footnote{\url{https://irdap.readthedocs.io/}}. The images shown are in polarized intensity with the contribution of the central star removed\footnote{Polarimetric observations of discs in multiple stellar systems are challenging due to the contamination of the signal by nearby stars, see \cite{Weber+2023} for more details.}. The ALMA images displayed in the galleries below are "quality assurance" products that were downloaded from the ALMA archive \footnote{\url{https://almascience.eso.org/aq/}} and used as they are.

\begin{table}[h]
\caption{Multiple stellar systems of interest with protoplanetary discs.}
    \begin{adjustwidth}{-\extralength}{0cm}
%        \newcolumntype{C}{>{\centering\arraybackslash}X}
        \begin{tabularx}{\fulllength}{CCCCCC}
            \toprule
            \textbf{Name} &  \textbf{RA (hh:mm:ss)}    & \textbf{Dec (dd:mm:ss)}  & \textbf{Number of stars} &  \textbf{Type of disc(s)$^1$} &  \textbf{References} \\
            \midrule
\multirow[m]{1}{*}{HT Lup              } & 15:45:12.868         & -34:17:30.64         & 3                    & CSD                  & \cite{Ghez+1997, Andrews+2018}\\
            \midrule
\multirow[m]{1}{*}{S CrA               } & 19:01:08.597         & -36:57:19.90         & 2                    & CSD                  & \cite{Stapelfeldt+1997, Cazzoletti+2019}\\
            \midrule
\multirow[m]{1}{*}{UY Aur              } & 04:51:47.389         & +30:47:13.55         & 2                    & CSD                  & \cite{Close+1998, Tang+2014}\\
            \midrule
\multirow[m]{1}{*}{T Tau               } & 04:21:59.432         & +19:32:06.43         & 3                    & CSD,CBD              & \cite{WhiteGhez2001}\\
            \midrule
\multirow[m]{1}{*}{HD 100453           } & 11:33:05.577         & -54:19:28.55         & 2                    & CSD                  & \cite{Chen+2006, Wagner+2015}\\
            \midrule
\multirow[m]{1}{*}{UZ Tau              } & 04:32:43.022         & +25:52:30.90         & 4                    & CSD,CBD              & \cite{WhiteGhez2001, Jensen+1996}\\
            \midrule
\multirow[m]{1}{*}{GG Tau              } & 04:32:30.351         & +17:31:40.49         & 5                    & CSD,CTD              & \cite{Leinert+1993, Dutrey+1994, DiFolco+2014, Roddier+1996, Silber+2000, Beck+2012, Keppler+2020}\\
            \midrule
\multirow[m]{1}{*}{GW Ori              } & 05:29:08.393         & +11:52:12.67         & 3                    & CBD,CTD              & \cite{Mathieu+1991, Mathieu+1995, Berger+2011, Kraus+2020}\\
            \midrule
\multirow[m]{1}{*}{HD 142527           } & 15:56:41.888         & -42:19:23.25         & 2                    & CSD,CBD              & \cite{Fukagawa+2006, Biller+2012, Marino+2015, Avenhaus+2017}\\
            \midrule
\multirow[m]{1}{*}{HD 98800            } & 11:22:05.290         & -24:46:39.76         & 4                    & CBD                  & \cite{Torres+1995,Koerner+2000}\\
            \bottomrule
        \end{tabularx}
    \end{adjustwidth}
    \noindent{\footnotesize{Table~\ref{tab:systemscomplete} in Appendix \ref{sec:systems} presents a more complete sample of multiple stellar systems with discs}.\\ 
    $^1$ CSD : CircumStellar Disc, CBD : CircumBinary Disc, CTD : CircumTriple Disc} 
    
\label{tab:systems}
\end{table}

\textbf{HT~Lup.} Located at a distance of $154\pm2$ pc \citep{Gaia+2016}, the HT~Lup system comprises three stars, all detected in millimetre wavelengths through both continuum and CO emission. HT~Lup~AB forms a close binary with a separation of $\sim0.1^{\prime\prime}$, while HT~Lup~C is positioned $\sim3^{\prime\prime}$ to the west, making the system a hierarchical triple. The individual discs around HT~Lup A and HT~Lup~B were first resolved by the DSHARP survey \citep{Andrews+2018,Kurtovic+2018}. The primary disc is relatively compact ($r\sim30$ au), while the secondary disc is only marginally resolved, suggesting truncation of both discs due to gravitational interactions. Interestingly, if the projected separation of HT~Lup~AB is assumed to represent the semi-major axis of a coplanar binary with an eccentricity of $0.2$, the expected truncation radius of the primary disc would be just $16$~au \citep{MirandaLai2015}. This discrepancy, along with the absence of material connecting the two discs, implies that the true separation of HT Lup~A and HT Lup~B is likely larger along the line of sight \citep{Kurtovic+2018}. Despite this, distinct spiral features in the continuum emission of HT~Lup~A suggest recent tidal interactions. Furthermore, CO intensity maps reveal that the primary and secondary discs are counter-rotating \citep{Kurtovic+2018}, suggesting the possibility of a retrograde orbit for HT Lup~AB.

\textbf{S~CrA.} S~CrA is a binary system with component masses of $0.70$~$M_{\odot}$ and $0.45$~$M_{\odot}$ \citep{Sullivan+2019}, separated by $\sim1^{\prime\prime}$. The individual circumstellar discs have been imaged in the near-infrared (NIR) \citep{Zhang+2023}, in continuum emission \citep{Cazzoletti+2019}, and in CO lines \citep{Gupta+2023}. In the continuum, the discs appear compact, while NIR observations reveal material connecting the two discs and extending into their surrounding environment. Recent analysis of NIR observations by \cite{Zhang+2023} identified spiral features in the primary disc, along with a tentative ring structure. Using astrometric measurements, the authors suggested an orbital inclination of $i=25\pm12^\circ$, which aligns with the plane of the inner disc but not with the inclination of $\sim57^\circ$ determined for the outer disc \citep{Gravity+2017}. While the observed spiral arms may result from tidal interactions with the secondary companion, alternative explanations include interactions with surrounding or infalling gas. These processes could also account for the diffuse emission observed in the environment of the binary system \citep{Gupta+2023,Gupta+2024}.

\textbf{UY~Aur.} Located in the Taurus–Aurigae star-forming region, UY~Aur consists of at least two stars of mass $0.6$~$M_{\odot}$ and $0.34$~$M_{\odot}$ separated by $0.88^{\prime\prime}$  \citep{HartiganKenyon2003}. A potential close companion to UY~Aur~B has been proposed to explain its photometric variability \citep{Tang+2014}. Early scattered-light and gas emission images of the system reveal extended circumbinary material, while millimetre emission clearly traces the individual discs associated with each star \citep{Hioki+2007, Tang+2014}. However, the rotation pattern observed in the circumbinary gas emission deviates from Keplerian motion, complicating the identification of a rotationally supported circumbinary disc \citep{Rota+2022}. The dust discs in UY~Aur are among the most compact in the Taurus region, likely due to significant tidal truncation \citep{Long+2019, Manara+2019}. Scattered-light and gas emission in the surrounding environment show numerous arc-like structures, which may result from interactions between the stars, the discs, and the out-flowing material \citep{Pyo+2014, Garufi+2024}. Further constraints on the orbit of UY~Aur~AB are necessary to robustly link these observed structures to the binary’s dynamics and better understand the system’s evolution.

\textbf{T~Tau.} T~Tau is one of the brightest optical objects in the Taurus cloud and is a triple stellar system. It consists of a $2.2$~$M_{\odot}$ star (T~Tau~N) orbiting a close binary (T~Tau~S) with component masses of $2.1$~$M_{\odot}$ and $0.4$~$M_{\odot}$ \citep{HerczegHillenbrand2014, Schaefer+2020}. High-resolution continuum imaging detected emission associated with each of the three stars \cite{Yamaguchi+2021}. T~Tau~N harbours a disc of $\sim22$~au, which features an apparent gap, while T~Tau~Sa and T~Tau~Sb host very compact discs. The orbit of the T~Tau~S binary has been well characterized, and its orbital period correlates strongly with arc-like structures observed to the south of the system \citep{Schaefer+2020, Beck+2020}. However, the kinematics of the T~Tau system are complex, likely reflecting interactions between the stars and surrounding infalling and out-flowing material \citep{Momose+1996, Beck+2020}. Based on modelling of 1.3~mm continuum images, the presence of a circumbinary disc (CBD) around T~Tau~S was proposed \cite{Beck+2020}. Notably, neither the individual discs nor the potential CBD associated with T~Tau~S appear to be aligned with the binary’s orbital plane \citep{Schaefer+2020, Beck+2020, Yamaguchi+2021}.

\textbf{HD~100453.} Located in the Lower Centaurus Association \citep{Kouwenhoven+2005}, HD~100453 is a Herbig star with a mass of $1.7$~$M_{\odot}$ and a nearby $0.2$~$M_{\odot}$ companion situated $\sim1^{\prime\prime}$ to the south-east \citep{Chen+2006}. The disc has been observed in near-infrared (NIR), millimetre continuum emission, and gas tracers \citep{Benisty+2017, vanderPlas+2019, Rosotti+2020}. These observations reveal a large cavity in the dust emission and an azimuthal asymmetry in the north-east, interpreted as the dust counterpart of extended spiral arms seen in the NIR and CO emission. HD~100453 has been extensively modelled to explain the origin of its spiral structures \citep{Dong+2016, Min+2017, Benisty+2017, Wagner+2018, vanderPlas+2019, Rosotti+2020, Gonzalez+2020}. The favoured interpretation attributes the spirals to tidal interactions with the companion, HD~100453~B. Both astrometric fitting and forward modelling of the disc structures suggest a misaligned orbit between the binary and the disc, which likely causes the disc to precess \citep{Gonzalez+2020}. Additionally, shadows observed in the NIR may indicate the presence of a misaligned inner disc. This scenario suggests that an undetected inner companion could have carved the cavity, dynamically decoupling the inner and outer discs \citep{Nealon+2020b}.

%%%%%%%%%%%%%%%%%%%%%%%%%%%%%%%%%%%%%%%%%%
\section{Circumbinary discs (CBDs) in multiple stellar systems}
\label{sec:CBDs}

Circumbinary discs (CBDs) in binary and multiple stellar systems represent complex and dynamic environments where interactions between the central binary and the disc strongly influence its structure and evolution. Unlike CSDs, which are directly associated with individual stars, CBDs encircle the entire binary system, and their morphology is shaped by the combined gravitational potential of the central stars. These discs are critical for understanding angular momentum transfer, accretion processes, and the formation of planets in binary systems.

\subsection{Size and shape of the inner CBD cavity}
\label{subsec:CBD_cavity}

One of the most distinctive features of a CBD is its large inner cavity, which forms due to tidal torques exerted by the binary. These torques prevent material from accreting directly onto the central stars, resulting in a cleared region whose size depends on the binary’s orbital separation, mass ratio, and eccentricity. For circular binaries, the cavity radius is typically $2-3$ times the binary separation, while eccentric binaries carve larger cavities, with sizes scaling approximately with $(1+e)$ \citep{ArtymowiczLubow1994}.

The cavity shape in circumbinary discs is determined by the binary's orbital parameters. For instance, CBDs with inner binaries with mass ratios high enough ($q=M_2/M_1$ above $0.05$) develop eccentric disc cavities --- both for circular and eccentric binaries \citep{Thun+2017, Hirsh+2020, Ragusa+2020, Penzlin+2024}. This mechanism eventually leads to the formation of lopsided (horseshoe-shaped) structures. These azimuthal asymmetries can translate into density enhancements, which are observable in gas tracers and dust continuum emission \citep[e.g.,][]{Ragusa+2021, Calcino+2023, Calcino+2024}. The dynamics within the cavity itself are complex, with material often flowing toward the stars via streamers that bridge the cavity \citep{ArtymowiczLubow1996, Duffell+2020, Ceppi+2022}. These accretion flows are essential for sustaining circumstellar material in binary systems, influencing accretion rates onto the stars and modulating binary evolution.

\subsection{Structure formation in CBDs}
\label{subsec:CBD_structures}

Gravitational torques exerted by the binary also drive the formation of structures such as spiral density waves and large-scale azimuthal asymmetries in CBDs. Unlike spirals in circumstellar discs, which are driven by external companions, spirals in CBDs are launched by the central binary itself. These features are most prominent in gas tracers, where density waves propagate outward from Lindblad resonances within the disc. The morphology of these spirals is influenced by the binary’s orbital parameters, including mass ratio and eccentricity, as well as the disc’s thermodynamic properties like viscosity and cooling efficiency \citep{Thun+2017, Price+2018, Penzlin+2024}.

Azimuthal asymmetries in the disc, such as horseshoe-shaped dust accumulations or lopsided gas structures, may arise from gravitational resonances or pressure traps at the inner cavity edge \citep{Ragusa+2017, Poblete+2019}. These features are observable in dust thermal emission and scattered light and may serve as sites for planetesimal growth. Additionally, simulations suggest that these structures may be transient or long-lived, depending on the binary’s stability and the disc’s physical properties \citep{Ragusa+2020}.

\subsection{CBD alignment}
\label{subsec:CBD_alignment}

The alignment of a CBD with the orbital plane of the binary depends on the binary’s inclination, mass ratio, and eccentricity, as well as the disc’s viscosity and angular momentum. Misaligned CBDs are often observed, particularly in systems where the binary orbits are themselves inclined relative to the total angular momentum vector of the system (see examples in Sect.~\ref{subsec:CBD_systems}). This misalignment can induce warps or twists in the disc, with different regions of the disc precessing at different rates due to differential torques \citep{Larwood+1996}.

For small misalignments, circumbinary discs (CBDs) can align with the binary’s orbital plane over time through viscous dissipation \citep{Facchini+2013}. However, if the initial inclination exceeds a critical angle, the disc may enter more complex dynamical regimes. In such cases, the disc can break into multiple misaligned rings or undergo Kozai-Lidov oscillations, where the disc’s inclination and eccentricity oscillate periodically due to the gravitational influence of the binary \citep{Facchini+2013, Aly+2015, Young+2023}. These effects are especially pronounced in systems with high binary eccentricities or weakly viscous discs.

For larger initial misalignments around eccentric binaries, CBDs may also achieve a state of polar alignment, where the disc becomes orthogonal to the binary’s orbital plane \citep{FaragoLaskar2010}. This occurs when the binary’s eccentricity-driven torques dominate over the viscous forces acting to align the disc with the orbital plane. In this regime, the CBD precesses stably around the binary’s eccentricity vector, while being almost in an orthogonal orientation with respect to the binary orbital plane \citep{Aly+2015, MartinLubow2017, CuelloGiuppone2019}. It is worth noting that several CBDs are observed having in an intermediate configuration, neither coplanar nor polar, and that these discs are found most of the time in high-order multiple systems, suggesting additional physics \citep{Czekala+2019, Ceppi+2023, Alaguero+2024}.

CBD misalignment also has significant implications for accretion. In misaligned systems, accretion streams may deliver material to circumstellar discs or directly onto the stars in a non-axisymmetric manner, potentially modulating accretion rates and driving variability in observed emission \citep{Franchini+2019, Duffell+2020, Ceppi+2022}. The resulting luminosity variability underscores the role of CBDs in shaping both stellar evolution and the potential for planet formation in these complex environments.

\subsection{CBDs of interest}
\label{subsec:CBD_systems}

\begin{figure}[H]
\centering
\begin{center}
\includegraphics[width=0.7\textwidth]{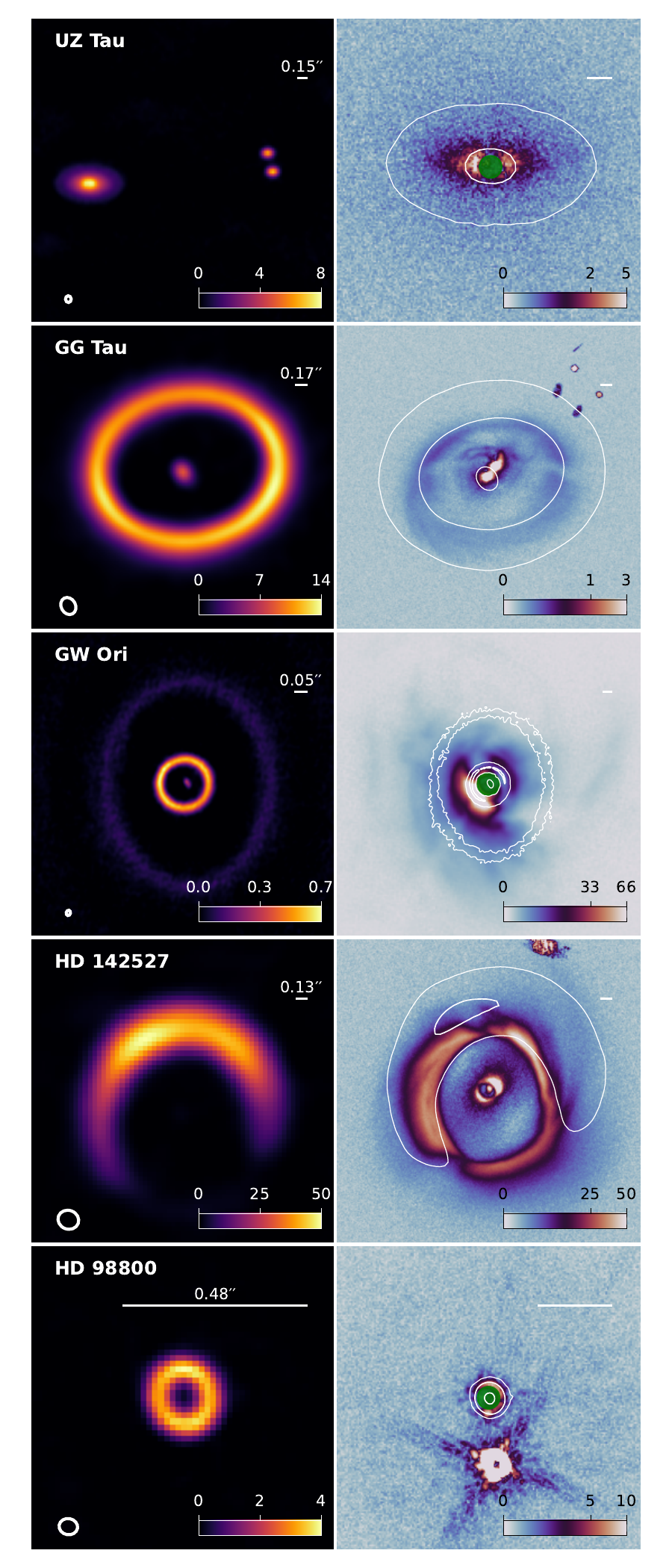}
\end{center}
\caption{Gallery of circumbinary discs (CBDs) with internal companions, processed using the same procedure as described in Figure \ref{fig:CSDs}. \mdy{The SPHERE data are taken at $1.65$ µm, except HD 142527 which was observed at $2.15$ µm. The colour scales have been adjusted to enhance the visibility of circumstellar extended emission.}}
\label{fig:CBDs}
\end{figure}

In the following, we describe ALMA and SPHERE observations of multiple stellar systems hosting circumbinary discs. We discuss the observational features of these systems and their interpretation in light of modelling efforts carried out by the community. These systems are listed in Table~\ref{tab:systems} and we show a gallery of recent observations in Figure \ref{fig:CBDs}.

\textbf{UZ~Tau.} The young stellar system UZ~Tau comprises four stars. UZ~Tau~E is a spectroscopic binary with a total mass of $1.3$ $M_{\odot}$, located $3.54^{\prime\prime}$ away from UZ~Tau~W. The latter consists of two stars separated by $0.37^{\prime\prime}$, with poorly constrained properties \citep{WhiteGhez2001, Prato+2002}. ALMA observations reveal a plain circumbinary disc (CBD) around UZ~Tau~E and two marginally resolved circumstellar discs (CSDs) associated with the stars in UZ~Tau~W. The sizes of the individual discs in UZ~Tau~W are consistent with tidal truncation. Moreover, their alignment with each other and the CBD suggests orbital coplanarity \citep{Garufi+2024}. Given the large observed separation between UZ~Tau~E and W, recent interactions between these components are unlikely. The CBD of UZ~Tau~E appears coplanar with its inner binary, which is consistent with the binary’s moderate eccentricity of $e=0.3$ \citep{Prato+2002, Jensen+2007}. However, contrary to theoretical predictions of a visible cavity of $\sim 0.3^{\prime\prime}$ \citep{ArtymowiczLubow1994}, the CBD shows only shows a small inner cavity of $\lesssim 0.08^{\prime\prime}$ \citep{Jennings+2022}, raising questions about its formation.

\textbf{GG~Tau.} GG~Tau is a hierarchical quintuple stellar system, consisting of GG~Tau~A (three stars) and GG~Tau~B (two stars) separated by $\sim 10^{\prime\prime}$ \citep{Leinert+1993}. GG~Tau~A, the more massive subsystem, includes GG~Tau~Aa and GG~Tau~Ab with estimated masses of $0.59$~$M_{\odot}$ and $\sim0.6+0.2$~$M_{\odot}$, respectively, and well-characterized internal orbits \citep{Phuong+2020, Duchene+2024}. This subsystem hosts a bright and long-studied circumtriple disc (CTD) which surrounds GG~Tau~A, featuring a prominent central cavity carved by the three stars (often considered as a binary) \citep[e.g.,][]{Dutrey+1994, Guilloteau+1999, Silber+2000, Duchene+2004}. High-resolution observations reveal asymmetries in the disc, including shadows along the RA axis and material within the cavity, possibly infalling toward the stars \citep{Keppler+2020}. Millimetre-wavelength observations also trace the circumstellar disc of GG~Tau~Aa \citep{Phuong+2020}. CO emission maps display prominent spiral arms extending into the cavity, which may result from gravitational interactions with GG~Tau~B or the inner stars of GG~Tau~A \citep{Phuong+2020,Pinte+2023}. The discs associated with GG~Tau~Ab and a potential CBD remain undetected, likely due to tidal truncation reducing their sizes. Modelling efforts suggest a misaligned configuration between the CBD and GG~Tau~A, which could explain some of the observed features \citep{Toci+2024}.

\textbf{GW~Ori.} The triple system GW~Ori, with its three close stars, was first resolved by \cite{Berger+2011}, with orbital parameters later refined by \cite{Czekala+2017} and \cite{Prato+2018}. The inner binary, GW~Ori~AB, has a period of $242$ days and a semi-major axis of $\sim 1$ au, while GW~Ori~AB-C orbits with a period of approximately $8$ years and a semi-major axis of $\sim 8$ au. Interestingly, the two orbital planes are misaligned by $\sim 14^{\circ}$ \citep{Kraus+2020}. The GW~Ori disc comprises three distinct rings, as revealed by scattered light and continuum observations \citep{Bi+2020, Kraus+2020}. The central regions are cleared of material by the stellar components, and all three rings are misaligned relative to the orbital planes of the stars, with the innermost ring further misaligned with the two outer rings. Modelling indicates that the observed disc tearing is driven by the gravitational torque from the inner triple star \citep{Kraus+2020, Young+2023, Rabago+2024}. Despite significant progress in characterising this system, many questions remain about the specific physical processes that led to the current configuration.

\textbf{HD~142527.} The disc of HD~142527 features a large inner cavity observable in both NIR and millimetre wavelengths \citep{Casassus+2015}. NIR observations also reveal central emission within the cavity, tracing material near the binary and casting shadows on the outer disc \citep{Biller+2012, Marino+2015, Avenhaus+2017}. In addition to the cavity, the disc exhibits striking features likely caused by disc-binary interactions. These include large spiral arms near the cavity edge and in the outer disc \citep{Fukagawa+2006, Christiaens+2014, Avenhaus+2017}, a bright horseshoe-shaped asymmetry in the northern disc \citep{Casassus+2015}, and fast flows connecting the inner and outer regions \citep{Casassus+2013}. Initial models suggested a binary orbit with a semi-major axis $a \lesssim 50$ au could account for these structures \citep{Price+2018}. However, more recent measurements place the binary at $a \sim 11$~au, insufficient to explain the observed features \citep{Nowak+2024}. These discrepancies suggest the presence of additional physics or a hidden companion influencing the system \citep{Penzlin+2024}. Notably, the new orbital solution reveals a mutual inclination of $\sim 8^{\circ}$ between the binary and the outer disc. 

\textbf{HD~98800.} Located $45$~pc away in the TW Hya association, HD~98800 is a $\sim10$ Myr-old quadruple system \citep{Soderblom+1996, Prato+2001, Torres+2008}. It consists of two close binaries, HD 98800~A and B, situated to the north and south of the system, respectively, with well-characterized orbits \citep{Zuniga-Fernandez+2021}. A small, tidally truncated disc around HD~98800~B has been resolved at $0.88$ mm, with a narrow ring width of $2$~au in dust emission and a central cavity in CO emission \citep{Koerner+2000, Akeson+2007, Andrews+2010}. These features align closely with models of tidal truncation by both the inner and outer binaries \citep{Ribas+2018, Kennedy+2019}. Remarkably, the plane of the CBD is in a polar configuration relative to the orbit of HD 98800~B, making it the first known system of its kind \citep{Kennedy+2019}. With such unique configurations now observed and modelled, ongoing theoretical efforts aim to explore how planet formation proceeds in polar discs \citep{Smallwood+2024b, Smallwood+2024c}.

%%%%%%%%%%%%%%%%%%%%%%%%%%%%%%%%%%%%%%%%%%
\section{Open questions in the field}
\label{sec:discussion}

As showcased in previous sections, the interaction between multiple stellar systems and protoplanetary discs offers a rich field of study. Over the past two decades, significant progress has been made in understanding how stellar multiplicity shapes disc structures and evolution. Here, we synthesise recent efforts made to address the connection between stellar orbits and disc morphology, the dynamics of dust multiple stellar systems, and the broad implications for planet formation.

\subsection{Linking stellar orbits to disc morphology}
\label{subsec:st_orb_to_discmorph}

The orbital characteristics of companion stars, including eccentricity, inclination, and semi-major axis, play a pivotal role in determining disc morphology. Eccentric orbits, for example, enhance tidal torques, creating large asymmetries within discs. In circumbinary discs (CBDs), circular and eccentric binaries can carve elongated cavities and excite strong spiral density waves \citep{ArtymowiczLubow1994, Ragusa+2020, Penzlin+2024}. Meanwhile, circumstellar discs (CSDs) in eccentric systems often exhibit truncated outer edges and steeper surface density profiles, which can accelerate radial drift and affect the distribution of solids. Inclination is another critical factor, particularly in systems where the companion star’s orbit is misaligned with the disc plane. Misaligned configurations induce precession and can lead to warped or even torn discs, as seen in simulations of systems like GW~Ori \citep{Kraus+2020, Rabago+2024}. These effects are not limited to binaries; in hierarchical triples, such as GG~Tau, interactions between the inner and outer stars further complicate the disc dynamics, potentially leading to distinctively warped or nested structures \citep{Phuong+2020, Duchene+2024, Toci+2024}.

\begin{figure}[H]
\begin{center}
\includegraphics[width=\textwidth]{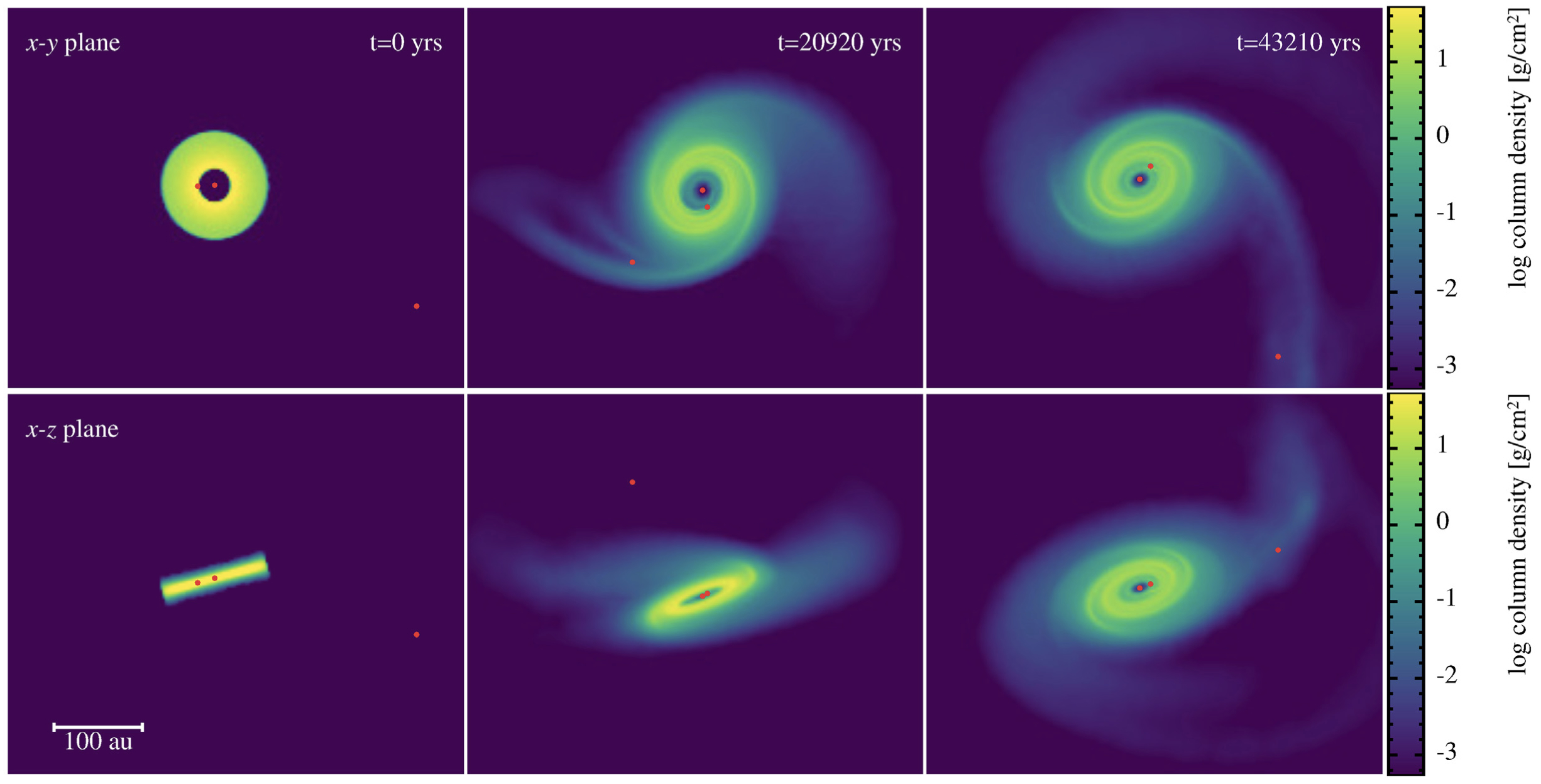}
\end{center}
\caption{Column density rendering of a hydrodynamical model of the system HD~100453 including a central star, a disc with an inner companion of $5$ M$_J$ at 20~au, and a misaligned external companion (star~B) with the following parameters: $q=0.12$, $a=207$~au, $e=0.32$, $i=49^\circ$, $\Omega=47^\circ$, $\omega=18^\circ$. The disc has an outer edge at 60~au and an aspect ratio $H/R$ equal to $0.05$. The top row shows the system in the plane of the sky, and the bottom row in a perpendicular plane. Stars and the planet are marked in red. Extended spirals and disc precession are driven by the external companion, while the inner planet carves a central cavity in the disc. Figure adapted from \citep{Nealon+2020b}, with permission of the authors.}
\label{fig:HD100453Nealon}
\end{figure}

Recent observational campaigns have provided compelling evidence linking stellar orbits to disc features. In systems like HD~142527, HD~100453, and AB~Aur, large-scale spiral arms and azimuthal asymmetries have been directly attributed to gravitational perturbations by stellar companions \citep{Price+2018, Gonzalez+2020, Poblete+2020}. However, in some cases, the binary scenario alone does not fully account for the observed complexity of disc structures, and in others, the proposed stellar companions remain below current detection limits. On top of this, the observed misalignments in systems such as GW~Ori further reinforce the connection between orbital dynamics and disc morphology. Figure~\ref{fig:HD100453Nealon} illustrates how hydrodynamical models of circumstellar discs replicate observed features such as spirals and precession while testing new dynamical scenarios. In this model, an inner gas giant planet (located at about 20~au) carves the inner disc cavity and efficiently drives the precession of the innermost regions, which are expected to cast shadows consistent with observations.

\begin{figure}[H]
\begin{center}
\includegraphics[width=\textwidth]{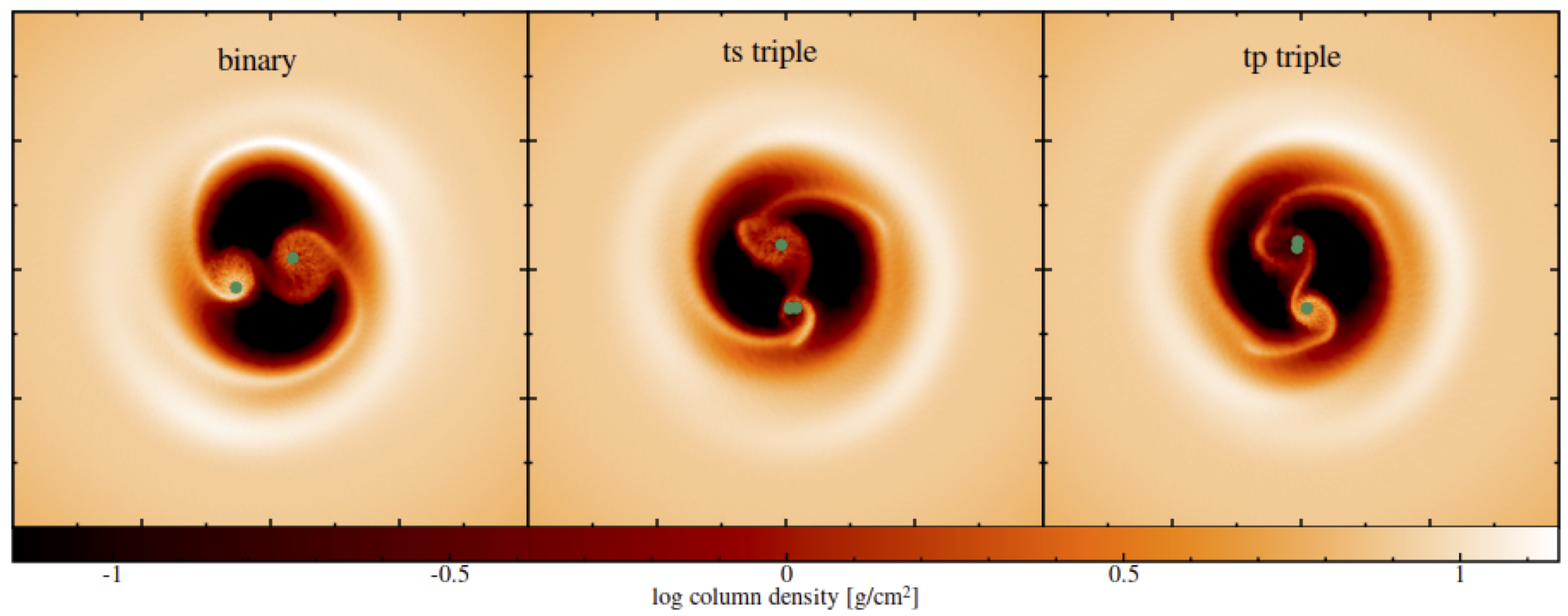}
\end{center}
\caption{Snapshots of three high-resolution simulations showing the gas density. Green dots represent the stars. The first column depicts the pure binary case, the central column shows a triple system obtained by splitting the secondary star of the binary (type ts), and the rightmost column shows a triple system obtained by splitting the primary star of the binary (type tp). These simulations evolved over 55 orbits of the outer binary. Figure adapted from \citep{Ceppi+2022}, with permission of the authors.}
\label{fig:Ceppi2022}
\end{figure}

Triple stellar systems introduce additional complexity, carving intricate inner cavities in their discs and giving rise to diverse morphologies and kinematics. Detailed mapping, as in GG~Tau, reveals how hierarchical arrangements shape accretion dynamics and structural features \citep{Keppler+2020, Pinte+2023}. Material crossing the cavity from the circumtriple disc to the inner stars is influenced by stellar motion, leading to luminosity variability, with accretion bursts exhibiting varying periodicity based on the stellar orbits \citep{Dunhill+2015, LaiMunoz2023}. As shown in Figure~\ref{fig:Ceppi2022}, recent hydrodynamical simulations demonstrate how stellar multiplicity affects accretion by comparing a pure binary case against two similar configurations of triple systems: one where the secondary star is split into two stars, and another where the primary star is split into two stars \cite{Ceppi+2022}. In principle, measuring accurate individual mass accretion rates onto the stars could provide additional constraints on the orbital parameters of young multiple stellar systems. At a more general level, multi-wavelength observations and long-term monitoring are essential for disentangling the roles of stellar components, reconstructing their 3D geometry, and refining parameters through modelling \citep{Kraus+2020, Smallwood+2021, Calcino+2023, Ragusa+2024, Calcino+2024}.

\subsection{Dust dynamics and growth in multiple stellar systems}
\label{subsec:dust_in_multi}

In multiple stellar systems, the dynamics of gas and dust are significantly influenced by the gravitational forces of nearby stellar companions. These tidal forces shape the overall structure of the protoplanetary disc, affecting the spatial distribution and evolution of gas and solids. Dust discs, in particular, are typically more compact than their gaseous counterparts due to the effects of radial drift \citep{Manara+2019, Rota+2022}. This process is particularly pronounced in CSDs, where the steep radial surface density profile of gas accelerates the inward drift of solids. As a result, dust grains are expected to be rapidly accreted onto the central star unless the local solid density becomes sufficiently high to trigger dust instabilities. One potential mechanism for halting this drift is the formation of self-induced dust traps \citep{Gonzalez+2017}, which can play a critical role in preventing dust loss in circumstellar discs within binary systems, thereby fostering favourable conditions to planet formation.

In CBDs, solids evolve differently: dust is expected to accumulate near the inner CBD edge, where tidal interactions with the binary companion create strong radial and azimuthal pressure gradients  \citep[e.g.][]{Ragusa+2017}. These pressure maxima serve as natural dust traps, facilitating the growth of larger solids \citep{Thun+2017, Chachan+2019, Coleman+2022}. Misaligned or eccentric CBDs introduce further complexity, as they are particularly efficient at capturing large dust particles. For instance, differences in the precession rates of gas and dust in misaligned CBDs can result in the formation of dust rings \cite{AlyLodato2020}, which have been recently extended to polar configurations \citep{Aly+2021, Longarini+2021, Smallwood+2024c}. Similarly, warped discs can induce oscillatory gas motions, creating high-density dust concentrations on dynamical timescales \citep{Aly+2024}. These dust trapping mechanisms, driven by gravitational perturbations, are critical for understanding the emergence of planets within CBDs, particularly in their inner regions. However, the intricate interplay of these processes remains to be fully explored.

The presence of nearby stellar companions also poses challenges for dust growth by exciting high collisional velocities between dust grains. At such velocities, collisions are more likely to result in fragmentation than in coagulation \cite{DominikDullemond2024}, potentially hindering the formation of larger particles. Yet, the very same perturbers imprint high-density structures in the disc, such as spirals, warps, and crescent-shaped asymmetries, as discussed in Sections~\ref{subsec:CSD_structures} and \ref{subsec:CBD_structures}. These features could serve as efficient dust traps, potentially fostering grain growth within localized regions. Despite advances in understanding, the competition between fragmentation and growth remains poorly constrained in multiple stellar systems. High-resolution observations, laboratory experiments, and sophisticated numerical simulations are required to disentangle the effects of perturbations from the intrinsic growth processes of dust particles.

In this context, the advent of ALMA has revolutionised the study of dust properties in protoplanetary discs, offering unprecedented resolution and sensitivity. A widely used method for characterizing grain properties involves fitting a dust model to the system’s spectral energy distribution (SED), enabling insights into grain size distributions and compositions regardless of the stellar multiplicity \citep[e.g.][]{Ribas+2020}. Scattered light imaging in the near-infrared (NIR) provides complementary constraints on grain properties, but its applicability is limited in discs with non-axisymmetric structures \citep[e.g.][]{Ginski+2023}. In the millimetre regime, fitting the SED slope across multiple wavelengths has emerged as a robust technique for probing grain size variations within discs. Resolved observations, often radially or azimuthally averaged to reduce computational costs, allow for the spatial modelling of grain properties \citep[e.g.][]{Macias+2021, Guidi+2022}. Systematic studies applying these methods have begun to characterise dust populations in multiple stellar systems \citep{Jiang+2024, Sierra+2024}. Intriguingly, \cite{Sierra+2024} found no correlation between the presence of multiple stars and the maximum grain size in their sample. Similarly, \cite{Riviere-Marichalar+2023} detected no signs of dust growth at the location of a continuum over-density in AB~Aur, while \cite{Jiang+2024} concluded that grain growth in the outer regions of discs is often constrained by the fragmentation barrier \citep{Birnstiel+2011}. Remarkably, new methods are emerging to better characterise azimuthal asymmetries in protoplanetary discs \citep[e.g.][]{Ribas+2024}, which are highly relevant for multiple stellar systems \citep[e.g.][]{Alaguero+2024}. Overall, a larger sample is required to be able to assess the influence of stellar multiplicity on dust growth.

Numerical simulations offer critical insights into the growth and evolution of dust grains, though they often come at the expense of high spatial resolution. For example, 1D simulations can accurately model multiple dust populations, incorporating collisions between grains of varying sizes and porosity evolution \citep[e.g.][]{StammlerBirnstiel2022}. However, the introduction of multiple stars breaks the symmetry of the problem, necessitating a multi-dimensional approach. Recent advances in 3D simulations have enabled the inclusion of dust size and porosity evolution in single stellar systems, though inter-grain collisions are typically limited to particles of similar sizes \citep{Vericel+2021, Michoulier+2024}. These simulations have shown that particle growth to decimetre scales requires porosity and compaction over time, typically within the snowline, where fragmentation velocities are higher. In multiple stellar systems, 3D hydrodynamical simulations with live radiative transfer indicate that snowlines tend to align with the cavity edges of CBDs \citep{Clanton2013, PierensNelson2024}. This alignment implies that significant dust growth is constrained to the inner regions of CBDs where conditions are more favourable for planetesimal formation.

\subsection{Implications for planet formation}
\label{subsec:planetform_in_multi}

The modified evolutionary pathways of discs in multiple stellar systems, eventually impact the formation of planetesimals and hence the resulting planetary architectures. In such systems, disc truncation by stellar companions limits the available mass for planet formation while concentrating material in smaller, high-density regions. For instance, the concentration of solids near the cavity edges of CSDs and CBDs could explain the population of short-period exoplanets in binaries (\url{https://exoplanet.eu/planets_binary/}). If confirmed, this trend would reinforce the role of stellar multiplicity in shaping exoplanet diversity. On top of this, the presence of strong tidal interactions in multiple systems could also influence planetary migration pathways \citep[e.g.][]{Gianuzzi+2023}. This formation scenario may favour the formation of rocky, terrestrial planets over gas giants, contrasting with the broader diversity of planetary types seen in single stellar systems \citep[e.g.][]{CuelloSucerquia2024}.

Moreover, misaligned discs constitute unique planetary factories for producing unconventional orbital configurations. In fact, simulations and observations suggest that such systems, especially polar CBDs, could host planets on highly inclined, eccentric, or even retrograde orbits \citep[e.g.][]{DoolinBlundell2011, CuelloGiuppone2019, Chen+2023}. These planetary outcomes reflect the early gravitational influence of stellar companions on the disc and highlight how multiplicity drives diversity in exoplanetary architectures. For example, the inclined planetary orbits observed in systems like TOI-1338/BEBOP-1 (b and c) serve as valuable test beds for studying planet formation around binary stars \citep{Standing+2023}. In the near future, the continued progress of surveys is expected to reveal additional multi-planetary architectures.

However, the structural and dynamical complexity of discs in multiple stellar systems poses significant challenges for detecting and characterising exoplanets. Techniques such as astrometry, transits, and radial velocity suffer from noise and systematic effects introduced by stellar companions. Despite these challenges, the emerging understanding of disc dynamics and their influence on planet formation can directly inform observational strategies. For example, high-precision astrometric missions like \textit{Gaia} \cite{Perryman+2014, Gaia+2016, Gaia+2023} and advanced radial velocity surveys like BEBOP \cite{Martin+2019} now offer improved sensitivity to planets in these complex environments. The upcoming search will be crucial to better understand the planet occurrence rate and the differences with respect to single stellar systems
\citep{MarzariThebault2019, BonavitaDesidera2020, MoeKratter2021}.

%%%%%%%%%%%%%%%%%%%%%%%%%%%%%%%%%%%%%%%%%%
\section{\mdy{Discussion and conclusions}}
\label{sec:conclusions}

In this review, we have highlighted the profound impact of stellar multiplicity on the dynamics, structure, and evolution of protoplanetary discs --- demonstrating how their interplay offers unique insights into the complex processes of planet formation. We summarise our main conclusions and future developments as follows:

\begin{itemize}
    \item \textbf{Stellar multiplicity shapes disc structures and planet formation.} The presence of multiple stars significantly alters the dynamics and morphology of protoplanetary discs, leading to truncation, misalignment, and dust traps. These structures influence where and how planetesimals form, potentially favouring rocky planets due to limited mass reservoirs for gas giant formation. 
    
    \item \textbf{Misaligned and polar discs produce unique planetary architectures.} Misaligned and polar circumbinary discs drive the formation of planets with unconventional orbital configurations, such as inclined, eccentric, or retrograde orbits. These configurations exemplify the diversity of planetary outcomes shaped by stellar multiplicity.

    \item \textbf{The observation of exoplanets in multiple systems remains challenging,} mainly due to noise and systematic effects from stellar companions. This is one of the current bottlenecks in the field. However, advances in astrometry (e.g., \textit{Gaia} \cite{Gaia+2016}), radial velocity surveys (e.g., BEBOP \cite{Martin+2019}), and high-resolution imaging are beginning to address these challenges, enabling exploration of this under-sampled population.
\end{itemize}

Future research on stellar multiplicity and its influence on planet formation will require a coordinated effort combining high-resolution observations with advanced numerical modelling. Instruments such as the James Webb Space Telescope (JWST) and ALMA together with the next generation of Extremely Large Telescopes (ELTs) and Square Kilometre Array (SKA) will provide critical constraints regarding the structure and dynamics of protoplanetary discs, particularly in their inner regions where planets actively form. When paired with 3D hydrodynamical simulations incorporating realistic treatments of dust growth, fragmentation, and companion-induced perturbations, these efforts will deepen our understanding of planet formation in multiple stellar systems.

Expanding the observational sample of discs in multiple stellar systems (see our extended list in Appendix~\ref{sec:systems}) will also help establish statistical links between stellar configurations, disc properties, and emerging planetary architectures. This will clarify the role of stellar multiplicity in shaping planetary diversity, particularly in systems with polar or misaligned discs, which may offer unique pathways for planet formation. These efforts will guide the development of targeted strategies to detect and characterise exoplanets in multiple stellar systems --- a crucial step toward understanding the full spectrum of planetary outcomes across all stellar environments.

To conclude on a historical note, we would like to quote Sir William Herschel, a pioneering figure in astronomy and the study of multiple stellar systems, who in 1789 wrote: \textit{"This method of viewing the heavens seems to throw them into a new kind of light. They now are seen to resemble a luxuriant garden, which contains the greatest variety of productions, in different flourishing beds; and one advantage we may at least reap from it is, that we can, as it were, extend the range of our experience to an immense duration"}.

%This section is not mandatory, but can be added to the manuscript if the discussion is unusually long or complex.

%%%%%%%%%%%%%%%%%%%%%%%%%%%%%%%%%%%%%%%%%%
%\section{Patents}
%This section is not mandatory, but may be added if there are patents resulting from the work reported in this manuscript.

%%%%%%%%%%%%%%%%%%%%%%%%%%%%%%%%%%%%%%%%%%
\vspace{6pt} 

%%%%%%%%%%%%%%%%%%%%%%%%%%%%%%%%%%%%%%%%%%
%% optional
%\supplementary{The following supporting information can be downloaded at:  \linksupplementary{s1}, Figure S1: title; Table S1: title; Video S1: title.}


%\begin{tabularx}{\textwidth}{lll}
%\toprule
%\textbf{Name} & \textbf{Type} & \textbf{Description} \\
%\midrule
%S1 & Python script (.py) & Script of python source code used in XX \\
%S2 & Text (.txt) & Script of modelling code used to make Figure X \\
%S3 & Text (.txt) & Raw data from experiment X \\
%S4 & Video (.mp4) & Video demonstrating the hardware in use \\
%\bottomrule
%\end{tabularx}
%}

%%%%%%%%%%%%%%%%%%%%%%%%%%%%%%%%%%%%%%%%%%
\authorcontributions{Conceptualization, N.C., A.A, P.P.; methodology, N.C., A.A, P.P.; software, A.A.; data curation, A.A, N.C.; writing---original draft preparation, N.C., A.A, P.P.; writing---review and editing, N.C., A.A; visualization, A.A.; supervision, N.C.; project administration, N.C.; funding acquisition, N.C. All authors have read and agreed to the published version of the manuscript.}

\funding{This project has received funding from the European Research Council (ERC) under the European Union Horizon Europe programme (grant agreement No. 101042275, project Stellar-MADE).}

\institutionalreview{Not applicable.}

\dataavailability{
%Details regarding where data supporting reported results can be found, including links to publicly archived datasets analyzed or generated during the study.
The SPHERE observational data used in this study have been downloaded from the ESO Science Archive Facility using the following queries :
\begin{itemize}

    \item \href{http://archive.eso.org/wdb/wdb/eso/sphere/query?wdbo=html\%2fdisplay\&max_rows_returned=1000\&target=HT\%20Lup\&resolver=simbad\&coord_sys=eq\&coord1=\&coord2=\&box=00\%2000\%2030\&format=sexagesimal\&tab_wdb_input_file=on\&wdb_input_file=\&night=\&stime=\&starttime=12\&etime=\&endtime=12\&tab_prog_id=on\&prog_id=\&prog_type=\%25\&obs_mode=\%25\&pi_coi=\&pi_coi_name=PI_only\&prog_title=\&tab_dp_id=on\&dp_id=\&tab_exptime=on\&exptime=64\&tab_dp_cat=on\&dp_cat=\%25\&tab_dp_type=on\&dp_type=\%25\&dp_type_user=\&tab_dp_tech=on\&dp_tech=POLARIMETRY\%25\&dp_tech_user=\&tab_ob_container_id=on\&ob_container_id=\&ob_container_type=\%25\&tab_ob_container_parent_id=on\&ob_container_parent_id=\&ob_container_parent_type=\%25\&tab_ob_id=on\&ob_id=\&tab_obs_targ_name=on\&obs_targ_name=\&tpl_id=\&tab_tpl_start=on\&tpl_start=\&tpl_expno=\&tpl_nexp=\&obs_tplno=\&tab_seq_arm=on\&seq_arm=IRDIS\&tab_ins3_opti5_name=on\&ins3_opti5_name=\%25\&tab_ins3_opti6_name=on\&ins3_opti6_name=\%25\&tab_ins_comb_vcor=on\&ins_comb_vcor=\%25\&tab_ins_comb_iflt=on\&ins_comb_iflt=\%25\&tab_ins_comb_pola=on\&ins_comb_pola=\%25\&tab_ins_comb_icor=on\&ins_comb_icor=\%25\&tab_det_dit1=on\&det_dit1=\&tab_det_seq1_dit=on\&det_seq1_dit=\&tab_det_ndit=on\&det_ndit=\&tab_det_read_curname=on\&det_read_curname=\%25\&ins2_opti2_name=\%25\&tab_ins4_drot2_mode=on\&ins4_drot2_mode=\%25\&tab_ins1_filt_name=on\&ins1_filt_name=B_H\&tab_ins1_opti1_name=on\&ins1_opti1_name=\%25\&tab_ins1_opti2_name=on\&ins1_opti2_name=\%25\&tab_ins4_opti11_name=on\&ins4_opti11_name=\%25\&tab_fwhm_avg=on\&fwhm_avg=\&airmass_range=\&night_flag=\%25\&moon_illu=\&order=\&}{HT Lup},

    \item \href{http://archive.eso.org/wdb/wdb/eso/sphere/query?wdbo=html\%2fdisplay\&max_rows_returned=1000\&target=S\%20CrA\&resolver=simbad\&coord_sys=eq\&coord1=\&coord2=\&box=00\%2000\%2030\&format=sexagesimal\&tab_wdb_input_file=on\&wdb_input_file=\&night=\&stime=\&starttime=12\&etime=\&endtime=12\&tab_prog_id=on\&prog_id=\&prog_type=\%25\&obs_mode=\%25\&pi_coi=\&pi_coi_name=PI_only\&prog_title=\&tab_dp_id=on\&dp_id=\&tab_exptime=on\&exptime=8\&tab_dp_cat=on\&dp_cat=\%25\&tab_dp_type=on\&dp_type=\%25\&dp_type_user=\&tab_dp_tech=on\&dp_tech=POLARIMETRY\%25\&dp_tech_user=\&tab_ob_container_id=on\&ob_container_id=\&ob_container_type=\%25\&tab_ob_container_parent_id=on\&ob_container_parent_id=\&ob_container_parent_type=\%25\&tab_ob_id=on\&ob_id=\&tab_obs_targ_name=on\&obs_targ_name=\&tpl_id=\&tab_tpl_start=on\&tpl_start=\&tpl_expno=\&tpl_nexp=\&obs_tplno=\&tab_seq_arm=on\&seq_arm=IRDIS\&tab_ins3_opti5_name=on\&ins3_opti5_name=\%25\&tab_ins3_opti6_name=on\&ins3_opti6_name=\%25\&tab_ins_comb_vcor=on\&ins_comb_vcor=\%25\&tab_ins_comb_iflt=on\&ins_comb_iflt=\%25\&tab_ins_comb_pola=on\&ins_comb_pola=\%25\&tab_ins_comb_icor=on\&ins_comb_icor=\%25\&tab_det_dit1=on\&det_dit1=\&tab_det_seq1_dit=on\&det_seq1_dit=\&tab_det_ndit=on\&det_ndit=\&tab_det_read_curname=on\&det_read_curname=\%25\&ins2_opti2_name=\%25\&tab_ins4_drot2_mode=on\&ins4_drot2_mode=\%25\&tab_ins1_filt_name=on\&ins1_filt_name=B_H\&tab_ins1_opti1_name=on\&ins1_opti1_name=\%25\&tab_ins1_opti2_name=on\&ins1_opti2_name=\%25\&tab_ins4_opti11_name=on\&ins4_opti11_name=\%25\&tab_fwhm_avg=on\&fwhm_avg=\&airmass_range=\&night_flag=\%25\&moon_illu=\&order=\&}{S CrA},

    \item \href{http://archive.eso.org/wdb/wdb/eso/sphere/query?wdbo=html\%2fdisplay\&max_rows_returned=1000\&target=UY\%20Aur\&resolver=simbad\&coord_sys=eq\&coord1=\&coord2=\&box=00\%2000\%2030\&format=sexagesimal\&tab_wdb_input_file=on\&wdb_input_file=\&night=\&stime=\&starttime=12\&etime=\&endtime=12\&tab_prog_id=on\&prog_id=\&prog_type=\%25\&obs_mode=\%25\&pi_coi=\&pi_coi_name=PI_only\&prog_title=\&tab_dp_id=on\&dp_id=\&tab_exptime=on\&exptime=16\&tab_dp_cat=on\&dp_cat=\%25\&tab_dp_type=on\&dp_type=\%25\&dp_type_user=\&tab_dp_tech=on\&dp_tech=POLARIMETRY\%25\&dp_tech_user=\&tab_ob_container_id=on\&ob_container_id=\&ob_container_type=\%25\&tab_ob_container_parent_id=on\&ob_container_parent_id=\&ob_container_parent_type=\%25\&tab_ob_id=on\&ob_id=\&tab_obs_targ_name=on\&obs_targ_name=\&tpl_id=\&tab_tpl_start=on\&tpl_start=\&tpl_expno=\&tpl_nexp=\&obs_tplno=\&tab_seq_arm=on\&seq_arm=IRDIS\&tab_ins3_opti5_name=on\&ins3_opti5_name=\%25\&tab_ins3_opti6_name=on\&ins3_opti6_name=\%25\&tab_ins_comb_vcor=on\&ins_comb_vcor=\%25\&tab_ins_comb_iflt=on\&ins_comb_iflt=\%25\&tab_ins_comb_pola=on\&ins_comb_pola=\%25\&tab_ins_comb_icor=on\&ins_comb_icor=\%25\&tab_det_dit1=on\&det_dit1=\&tab_det_seq1_dit=on\&det_seq1_dit=\&tab_det_ndit=on\&det_ndit=\&tab_det_read_curname=on\&det_read_curname=\%25\&ins2_opti2_name=\%25\&tab_ins4_drot2_mode=on\&ins4_drot2_mode=\%25\&tab_ins1_filt_name=on\&ins1_filt_name=B_H\&tab_ins1_opti1_name=on\&ins1_opti1_name=\%25\&tab_ins1_opti2_name=on\&ins1_opti2_name=\%25\&tab_ins4_opti11_name=on\&ins4_opti11_name=\%25\&tab_fwhm_avg=on\&fwhm_avg=\&airmass_range=\&night_flag=\%25\&moon_illu=\&order=\&}{UY Aur},

     \item \href{http://archive.eso.org/wdb/wdb/eso/sphere/query?wdbo=html\%2fdisplay\&max_rows_returned=1000\&target=T\%20Tau\&resolver=simbad\&coord_sys=eq\&coord1=\&coord2=\&box=00\%2000\%2030\&format=sexagesimal\&tab_wdb_input_file=on\&wdb_input_file=\&night=\&stime=\&starttime=12\&etime=\&endtime=12\&tab_prog_id=on\&prog_id=\&prog_type=\%25\&obs_mode=\%25\&pi_coi=\&pi_coi_name=PI_only\&prog_title=\&tab_dp_id=on\&dp_id=\&tab_exptime=on\&exptime=16\&tab_dp_cat=on\&dp_cat=\%25\&tab_dp_type=on\&dp_type=\%25\&dp_type_user=\&tab_dp_tech=on\&dp_tech=POLARIMETRY\%25\&dp_tech_user=\&tab_ob_container_id=on\&ob_container_id=\&ob_container_type=\%25\&tab_ob_container_parent_id=on\&ob_container_parent_id=\&ob_container_parent_type=\%25\&tab_ob_id=on\&ob_id=\&tab_obs_targ_name=on\&obs_targ_name=\&tpl_id=\&tab_tpl_start=on\&tpl_start=\&tpl_expno=\&tpl_nexp=\&obs_tplno=\&tab_seq_arm=on\&seq_arm=IRDIS\&tab_ins3_opti5_name=on\&ins3_opti5_name=\%25\&tab_ins3_opti6_name=on\&ins3_opti6_name=\%25\&tab_ins_comb_vcor=on\&ins_comb_vcor=\%25\&tab_ins_comb_iflt=on\&ins_comb_iflt=\%25\&tab_ins_comb_pola=on\&ins_comb_pola=\%25\&tab_ins_comb_icor=on\&ins_comb_icor=\%25\&tab_det_dit1=on\&det_dit1=\&tab_det_seq1_dit=on\&det_seq1_dit=\&tab_det_ndit=on\&det_ndit=\&tab_det_read_curname=on\&det_read_curname=\%25\&ins2_opti2_name=\%25\&tab_ins4_drot2_mode=on\&ins4_drot2_mode=\%25\&tab_ins1_filt_name=on\&ins1_filt_name=\%25\&tab_ins1_opti1_name=on\&ins1_opti1_name=\%25\&tab_ins1_opti2_name=on\&ins1_opti2_name=\%25\&tab_ins4_opti11_name=on\&ins4_opti11_name=\%25\&tab_fwhm_avg=on\&fwhm_avg=\&airmass_range=\&night_flag=\%25\&moon_illu=\&order=\&}{T Tau},
    
    \item \href{http://archive.eso.org/wdb/wdb/eso/sphere/query?wdbo=html\%2fdisplay\&max_rows_returned=1000\&target=HD\%20100453\&resolver=simbad\&coord_sys=eq\&coord1=\&coord2=\&box=00\%2000\%2030\&format=sexagesimal\&tab_wdb_input_file=on\&wdb_input_file=\&night=\&stime=\&starttime=12\&etime=\&endtime=12\&tab_prog_id=on\&prog_id=\&prog_type=\%25\&obs_mode=\%25\&pi_coi=\&pi_coi_name=PI_only\&prog_title=\&tab_dp_id=on\&dp_id=\&tab_exptime=on\&exptime=32\&tab_dp_cat=on\&dp_cat=\%25\&tab_dp_type=on\&dp_type=\%25\&dp_type_user=\&tab_dp_tech=on\&dp_tech=POLARIMETRY\%25\&dp_tech_user=\&tab_ob_container_id=on\&ob_container_id=\&ob_container_type=\%25\&tab_ob_container_parent_id=on\&ob_container_parent_id=\&ob_container_parent_type=\%25\&tab_ob_id=on\&ob_id=\&tab_obs_targ_name=on\&obs_targ_name=\&tpl_id=\&tab_tpl_start=on\&tpl_start=\&tpl_expno=\&tpl_nexp=\&obs_tplno=\&tab_seq_arm=on\&seq_arm=IRDIS\&tab_ins3_opti5_name=on\&ins3_opti5_name=\%25\&tab_ins3_opti6_name=on\&ins3_opti6_name=\%25\&tab_ins_comb_vcor=on\&ins_comb_vcor=\%25\&tab_ins_comb_iflt=on\&ins_comb_iflt=\%25\&tab_ins_comb_pola=on\&ins_comb_pola=\%25\&tab_ins_comb_icor=on\&ins_comb_icor=\%25\&tab_det_dit1=on\&det_dit1=\&tab_det_seq1_dit=on\&det_seq1_dit=\&tab_det_ndit=on\&det_ndit=\&tab_det_read_curname=on\&det_read_curname=\%25\&ins2_opti2_name=\%25\&tab_ins4_drot2_mode=on\&ins4_drot2_mode=\%25\&tab_ins1_filt_name=on\&ins1_filt_name=B_Ks\&tab_ins1_opti1_name=on\&ins1_opti1_name=\%25\&tab_ins1_opti2_name=on\&ins1_opti2_name=\%25\&tab_ins4_opti11_name=on\&ins4_opti11_name=\%25\&tab_fwhm_avg=on\&fwhm_avg=\&airmass_range=\&night_flag=\%25\&moon_illu=\&order=\&}{HD 100453},

    \item \href{http://archive.eso.org/wdb/wdb/eso/sphere/query?wdbo=html\%2fdisplay\&max_rows_returned=1000\&target=UZ\%20Tau\&resolver=simbad\&coord_sys=eq\&coord1=\&coord2=\&box=00\%2000\%2030\&format=sexagesimal\&tab_wdb_input_file=on\&wdb_input_file=\&night=\&stime=\&starttime=12\&etime=\&endtime=12\&tab_prog_id=on\&prog_id=\&prog_type=\%25\&obs_mode=\%25\&pi_coi=\&pi_coi_name=PI_only\&prog_title=\&tab_dp_id=on\&dp_id=\&tab_exptime=on\&exptime=64\&tab_dp_cat=on\&dp_cat=\%25\&tab_dp_type=on\&dp_type=\%25\&dp_type_user=\&tab_dp_tech=on\&dp_tech=POLARIMETRY\%25\&dp_tech_user=\&tab_ob_container_id=on\&ob_container_id=\&ob_container_type=\%25\&tab_ob_container_parent_id=on\&ob_container_parent_id=\&ob_container_parent_type=\%25\&tab_ob_id=on\&ob_id=\&tab_obs_targ_name=on\&obs_targ_name=\&tpl_id=\&tab_tpl_start=on\&tpl_start=\&tpl_expno=\&tpl_nexp=\&obs_tplno=\&tab_seq_arm=on\&seq_arm=IRDIS\&tab_ins3_opti5_name=on\&ins3_opti5_name=\%25\&tab_ins3_opti6_name=on\&ins3_opti6_name=\%25\&tab_ins_comb_vcor=on\&ins_comb_vcor=\%25\&tab_ins_comb_iflt=on\&ins_comb_iflt=\%25\&tab_ins_comb_pola=on\&ins_comb_pola=\%25\&tab_ins_comb_icor=on\&ins_comb_icor=\%25\&tab_det_dit1=on\&det_dit1=\&tab_det_seq1_dit=on\&det_seq1_dit=\&tab_det_ndit=on\&det_ndit=\&tab_det_read_curname=on\&det_read_curname=\%25\&ins2_opti2_name=\%25\&tab_ins4_drot2_mode=on\&ins4_drot2_mode=\%25\&tab_ins1_filt_name=on\&ins1_filt_name=B_H\&tab_ins1_opti1_name=on\&ins1_opti1_name=\%25\&tab_ins1_opti2_name=on\&ins1_opti2_name=\%25\&tab_ins4_opti11_name=on\&ins4_opti11_name=\%25\&tab_fwhm_avg=on\&fwhm_avg=\&airmass_range=\&night_flag=\%25\&moon_illu=\&order=\&}{UZ Tau},

    \item \href{http://archive.eso.org/wdb/wdb/eso/sphere/query?wdbo=html\%2fdisplay\&max_rows_returned=1000\&target=GG\%20Tau\&resolver=simbad\&coord_sys=eq\&coord1=\&coord2=\&box=00\%2000\%2030\&format=sexagesimal\&tab_wdb_input_file=on\&wdb_input_file=\&night=\&stime=\&starttime=12\&etime=\&endtime=12\&tab_prog_id=on\&prog_id=\&prog_type=\%25\&obs_mode=\%25\&pi_coi=\&pi_coi_name=PI_only\&prog_title=\&tab_dp_id=on\&dp_id=\&tab_exptime=on\&exptime=4\&tab_dp_cat=on\&dp_cat=\%25\&tab_dp_type=on\&dp_type=\%25\&dp_type_user=\&tab_dp_tech=on\&dp_tech=\%25\&dp_tech_user=\&tab_ob_container_id=on\&ob_container_id=\&ob_container_type=\%25\&tab_ob_container_parent_id=on\&ob_container_parent_id=\&ob_container_parent_type=\%25\&tab_ob_id=on\&ob_id=\&tab_obs_targ_name=on\&obs_targ_name=\&tpl_id=\&tab_tpl_start=on\&tpl_start=\&tpl_expno=\&tpl_nexp=\&obs_tplno=\&tab_seq_arm=on\&seq_arm=IRDIS\&tab_ins3_opti5_name=on\&ins3_opti5_name=\%25\&tab_ins3_opti6_name=on\&ins3_opti6_name=\%25\&tab_ins_comb_vcor=on\&ins_comb_vcor=\%25\&tab_ins_comb_iflt=on\&ins_comb_iflt=\%25\&tab_ins_comb_pola=on\&ins_comb_pola=\%25\&tab_ins_comb_icor=on\&ins_comb_icor=\%25\&tab_det_dit1=on\&det_dit1=\&tab_det_seq1_dit=on\&det_seq1_dit=\&tab_det_ndit=on\&det_ndit=\&tab_det_read_curname=on\&det_read_curname=\%25\&ins2_opti2_name=\%25\&tab_ins4_drot2_mode=on\&ins4_drot2_mode=\%25\&tab_ins1_filt_name=on\&ins1_filt_name=B_H\&tab_ins1_opti1_name=on\&ins1_opti1_name=\%25\&tab_ins1_opti2_name=on\&ins1_opti2_name=\%25\&tab_ins4_opti11_name=on\&ins4_opti11_name=\%25\&tab_fwhm_avg=on\&fwhm_avg=\&airmass_range=\&night_flag=\%25\&moon_illu=\&order=\&}{GG Tau},

    \item \href{http://archive.eso.org/wdb/wdb/eso/sphere/query?wdbo=html\%2fdisplay\&max_rows_returned=1000\&target=GW\%20Ori\&resolver=simbad\&coord_sys=eq\&coord1=\&coord2=\&box=00\%2000\%2030\&format=sexagesimal\&tab_wdb_input_file=on\&wdb_input_file=\&night=\&stime=\&starttime=12\&etime=\&endtime=12\&tab_prog_id=on\&prog_id=\&prog_type=\%25\&obs_mode=\%25\&pi_coi=\&pi_coi_name=PI_only\&prog_title=\&tab_dp_id=on\&dp_id=\&tab_exptime=on\&exptime=16\&tab_dp_cat=on\&dp_cat=\%25\&tab_dp_type=on\&dp_type=\%25\&dp_type_user=\&tab_dp_tech=on\&dp_tech=POLARIMETRY\%25\&dp_tech_user=\&tab_ob_container_id=on\&ob_container_id=\&ob_container_type=\%25\&tab_ob_container_parent_id=on\&ob_container_parent_id=\&ob_container_parent_type=\%25\&tab_ob_id=on\&ob_id=\&tab_obs_targ_name=on\&obs_targ_name=\&tpl_id=\&tab_tpl_start=on\&tpl_start=\&tpl_expno=\&tpl_nexp=\&obs_tplno=\&tab_seq_arm=on\&seq_arm=IRDIS\&tab_ins3_opti5_name=on\&ins3_opti5_name=\%25\&tab_ins3_opti6_name=on\&ins3_opti6_name=\%25\&tab_ins_comb_vcor=on\&ins_comb_vcor=\%25\&tab_ins_comb_iflt=on\&ins_comb_iflt=\%25\&tab_ins_comb_pola=on\&ins_comb_pola=\%25\&tab_ins_comb_icor=on\&ins_comb_icor=\%25\&tab_det_dit1=on\&det_dit1=\&tab_det_seq1_dit=on\&det_seq1_dit=\&tab_det_ndit=on\&det_ndit=\&tab_det_read_curname=on\&det_read_curname=\%25\&ins2_opti2_name=\%25\&tab_ins4_drot2_mode=on\&ins4_drot2_mode=\%25\&tab_ins1_filt_name=on\&ins1_filt_name=B_H\&tab_ins1_opti1_name=on\&ins1_opti1_name=\%25\&tab_ins1_opti2_name=on\&ins1_opti2_name=\%25\&tab_ins4_opti11_name=on\&ins4_opti11_name=\%25\&tab_fwhm_avg=on\&fwhm_avg=\&airmass_range=\&night_flag=\%25\&moon_illu=\&order=\&}{GW Ori},

    \item \href{http://archive.eso.org/wdb/wdb/eso/sphere/query?wdbo=html\%2fdisplay\&max_rows_returned=1000\&target=HD\%20142527\&resolver=simbad\&coord_sys=eq\&coord1=\&coord2=\&box=00\%2000\%2030\&format=sexagesimal\&tab_wdb_input_file=on\&wdb_input_file=\&night=\&stime=\&starttime=12\&etime=\&endtime=12\&tab_prog_id=on\&prog_id=\&prog_type=\%25\&obs_mode=\%25\&pi_coi=\&pi_coi_name=PI_only\&prog_title=\&tab_dp_id=on\&dp_id=\&tab_exptime=on\&exptime=64\&tab_dp_cat=on\&dp_cat=\%25\&tab_dp_type=on\&dp_type=\%25\&dp_type_user=\&tab_dp_tech=on\&dp_tech=POLARIMETRY\%25\&dp_tech_user=\&tab_ob_container_id=on\&ob_container_id=\&ob_container_type=\%25\&tab_ob_container_parent_id=on\&ob_container_parent_id=\&ob_container_parent_type=\%25\&tab_ob_id=on\&ob_id=\&tab_obs_targ_name=on\&obs_targ_name=\&tpl_id=\&tab_tpl_start=on\&tpl_start=\&tpl_expno=\&tpl_nexp=\&obs_tplno=\&tab_seq_arm=on\&seq_arm=IRDIS\&tab_ins3_opti5_name=on\&ins3_opti5_name=\%25\&tab_ins3_opti6_name=on\&ins3_opti6_name=\%25\&tab_ins_comb_vcor=on\&ins_comb_vcor=\%25\&tab_ins_comb_iflt=on\&ins_comb_iflt=\%25\&tab_ins_comb_pola=on\&ins_comb_pola=\%25\&tab_ins_comb_icor=on\&ins_comb_icor=\%25\&tab_det_dit1=on\&det_dit1=\&tab_det_seq1_dit=on\&det_seq1_dit=\&tab_det_ndit=on\&det_ndit=\&tab_det_read_curname=on\&det_read_curname=\%25\&ins2_opti2_name=\%25\&tab_ins4_drot2_mode=on\&ins4_drot2_mode=\%25\&tab_ins1_filt_name=on\&ins1_filt_name=\%25\&tab_ins1_opti1_name=on\&ins1_opti1_name=\%25\&tab_ins1_opti2_name=on\&ins1_opti2_name=\%25\&tab_ins4_opti11_name=on\&ins4_opti11_name=\%25\&tab_fwhm_avg=on\&fwhm_avg=\&airmass_range=\&night_flag=\%25\&moon_illu=\&order=\&'}{HD 142527},

    \item \href{http://archive.eso.org/wdb/wdb/eso/sphere/query?wdbo=html\%2fdisplay\&max_rows_returned=1000\&target=HD\%2098800\&resolver=simbad\&coord_sys=eq\&coord1=\&coord2=\&box=00\%2000\%2030\&format=sexagesimal\&tab_wdb_input_file=on\&wdb_input_file=\&night=\&stime=\&starttime=12\&etime=\&endtime=12\&tab_prog_id=on\&prog_id=\&prog_type=\%25\&obs_mode=\%25\&pi_coi=\&pi_coi_name=PI_only\&prog_title=\&tab_dp_id=on\&dp_id=\&tab_exptime=on\&exptime=16\&tab_dp_cat=on\&dp_cat=\%25\&tab_dp_type=on\&dp_type_user=OBJECT\&tab_dp_tech=on\&dp_tech=POLARIMETRY\%25\&dp_tech_user=\&tab_ob_container_id=on\&ob_container_id=\&ob_container_type=\%25\&tab_ob_container_parent_id=on\&ob_container_parent_id=\&ob_container_parent_type=\%25\&tab_ob_id=on\&ob_id=\&tab_obs_targ_name=on\&obs_targ_name=\&tpl_id=\&tab_tpl_start=on\&tpl_start=\&tpl_expno=\&tpl_nexp=\&obs_tplno=\&tab_seq_arm=on\&seq_arm=IRDIS\&tab_ins3_opti5_name=on\&ins3_opti5_name=\%25\&tab_ins3_opti6_name=on\&ins3_opti6_name=\%25\&tab_ins_comb_vcor=on\&ins_comb_vcor=\%25\&tab_ins_comb_iflt=on\&ins_comb_iflt=\%25\&tab_ins_comb_pola=on\&ins_comb_pola=\%25\&tab_ins_comb_icor=on\&ins_comb_icor=N_ALC_YJH_S\&tab_det_dit1=on\&det_dit1=\&tab_det_seq1_dit=on\&det_seq1_dit=\&tab_det_ndit=on\&det_ndit=\&tab_det_read_curname=on\&det_read_curname=\%25\&ins2_opti2_name=\%25\&tab_ins4_drot2_mode=on\&ins4_drot2_mode=\%25\&tab_ins1_filt_name=on\&ins1_filt_name=B_H\&tab_ins1_opti1_name=on\&ins1_opti1_name=\%25\&tab_ins1_opti2_name=on\&ins1_opti2_name=\%25\&tab_ins4_opti11_name=on\&ins4_opti11_name=\%25\&tab_fwhm_avg=on\&fwhm_avg=\&airmass_range=\&night_flag=\%25\&moon_illu=\&order=\&}{HD 98800}.
\end{itemize}
These raw data were reduced and processed using the publicly available \href{https://irdap.readthedocs.io/en/latest/index.html}{{\sc IRDAP} pipeline}.

The ALMA data used in this work were downloaded from the ALMA archive using the available data products from the following project codes :
\begin{itemize}
    \item HT Lup: 2016.1.00484.L  \cite{Andrews+2018},
    \item S CrA: 2019.1.01792.S,
    \item UY Aur: 2016.1.01164.S,
    \item T Tau: 2019.1.00703.S,
    \item HD 100453: 2017.1.01678.S,
    \item UZ Tau: 2016.1.01164.S,
    \item GG Tau: 2018.1.00532.S,
    \item GW Ori: 2018.1.00813.S,
    \item HD 142527: 2015.1.01353.S,
    \item HD 98800: 2017.1.00350.S. 
\end{itemize}
}

\acknowledgments{This research has made use of the Astrophysics Data System, funded by NASA under Cooperative Agreement 80NSSC21M00561.
This paper makes use of the following ALMA data: 
ADS/JAO.ALMA\#2017.1.01678.S,
2019.1.00703.S,
2016.1.00484.L, 
2016.1.01164.S,
2016.1.01164.S,
2019.1.01792.S,
2015.1.01353.S,
2018.1.00532.S,
2018.1.00813.S,
and 2017.1.00350.S.
ALMA is a partnership of ESO (representing its member states), NSF (USA) and NINS (Japan), together with NRC (Canada), NSC and ASIAA (Taiwan), and KASI (Republic of Korea), in cooperation with the Republic of Chile. The Joint ALMA Observatory is operated by ESO, AUI/NRAO and NAOJ. We thank Martina Toscani, Ulko, and the entire Stellar-MADE team for their support during this project.
}
% In this section you can acknowledge any support given which is not covered by the author contribution or funding sections. This may include administrative and technical support, or donations in kind (e.g., materials used for experiments)

\conflictsofinterest{The authors declare no conflicts of interest. The funders had no role in the design of the study; in the collection, analyses, or interpretation of data; in the writing of the manuscript; or in the decision to publish the results.} 

%%%%%%%%%%%%%%%%%%%%%%%%%%%%%%%%%%%%%%
%% Optional

%% Only for journal Encyclopedia
%\entrylink{The Link to this entry published on the encyclopedia platform.}

\abbreviations{Abbreviations}{
The following abbreviations are used in this manuscript:\\

\noindent 
\begin{tabular}{@{}ll}
CBD & Circumbinary Disc\\
CSD & Circumstellar Disc\\
CTD & Circumtriple Disc\\
ALMA & Atacama Large Millimeter Array\\
SPHERE & Spectro-Polarimetric High-Contrast Exoplanet Research\\
VLA & Karl G. Jansky Very Large Array\\
VLT & Very Large Telescope\\
NIR & Near Infrared
\end{tabular}
}

%%%%%%%%%%%%%%%%%%%%%%%%%%%%%%%%%%%%%%%%%%
%% Optional
\appendixtitles{yes} % Leave argument "no" if all appendix headings stay EMPTY (then no dot is printed after "Appendix A"). If the appendix sections contain a heading then change the argument to "yes".
\appendixstart
\appendix

%\section[\appendixname~\thesection]{}
%\subsection[\appendixname~\thesubsection]{}

\section{Orbital parameters}
\label{sec:app_orb}

Here, we list the relevant orbital parameters, along with their definitions, which are key for describing binary orbits \citep{MurrayDermott1999}. These parameters are illustrated in Fig.~\ref{fig:orbital-elements}.
\begin{enumerate}
\item The semi-major axis ($a$), which represents the average distance between the star and the system’s centre of mass. It is an indicator of the orbit’s size and, by extension, the stellar orbital period around the common centre of mass. For a binary system with masses $m_1$ and $m_2$, this can be obtained as follows:
\begin{equation}
    T=2\pi\sqrt{\frac{a^3}{G\left(m_1+m_2\right)}},
\end{equation}
where $G$ is the gravitational constant. It is however common to work with the orbital frequency $\Omega_k=2\pi/T$.
\item The eccentricity ($e$) describes the shape of the orbit. A value of $0$ corresponds to a perfectly circular orbit, while values approaching $1$ indicate highly elliptical orbits. Eccentricity influences the variation in distance between the two bodies during their motion. The closest and farthest points along the orbit are known as the periapsis and apoapsis, respectively, given by $r_{\rm min}=a(1-e)$ and $r_{\rm max}=a(1+e)$. 
\item The inclination ($i$) measures the angle between the orbital plane and the reference plane, typically the plane of the sky. This parameter is critical for understanding the orbital orientation relative to the observer’s line of sight.
\item The longitude of the ascending node ($\Omega$) represents the point on the orbit where the star crosses the reference plane moving from South to North (towards the $\aries$ sign in Figure \ref{fig:orbital-elements}). This parameter is essential for determining the orbit’s orientation in space.
\item The argument of periapsis ($\omega$) specifies the orbital orientation of the ellipse within its plane. It defines the angular position of the periapsis relative to the ascending node.
\item The true anomaly ($\nu$) is the angle between the direction of the periapsis and the star’s current position. Alternatively, the mean anomaly ($M$) provides a convenient way to represent the orbital angle between the periapsis and the star’s position. It accounts for Kepler’s second law by defining angles such that each successive segment encloses equal areas within the elliptical orbit, corresponding to equal time intervals.
\end{enumerate}

\section{List of young multiple stellar systems with protoplanetary discs}
\label{sec:systems}

Below, we provide an extensive (although not exhaustive) list of multiple stellar systems of interest with protoplanetary discs. The latter can either be circumstellar (CSD), circumbinary (CBD), or circumtriple (CTD).

\setcounter{table}{0}
\renewcommand{\thetable}{B\arabic{table}}

\begin{adjustwidth}{-\extralength}{0cm}

\setlength{\LTleft}{-3cm}
\setlength{\LTright}{0cm}
\setlength{\LTright}{0cm}
\begin{longtable}{cccccc}
\caption{List of multiple stellar systems with protoplanetary discs.\label{tab:systemscomplete}}\\
            \toprule
            \textbf{Name} &  \textbf{RA (hh:mm:ss)}    & \textbf{Dec (dd:mm:ss)}  & \textbf{Number of stars} &  \textbf{Type of disc(s)}$^1$ &  \textbf{References} \\
            \midrule
            \endfirsthead
\caption{continued.}\\
            \toprule
            \textbf{Name} &  \textbf{RA (hh:mm:ss)}    & \textbf{Dec (dd:mm:ss)}  & \textbf{Number of stars} &  \textbf{Type of disc(s)} &  \textbf{References} \\
            \midrule
            \endhead
\multirow[m]{1}{*}{99 Her              } & 18:07:01.591         & +30:33:43.58         & 2                    & CBD                  & \cite{Kennedy+2012}\\
            \midrule
\multirow[m]{1}{*}{AK Sco              } & 16:54:44.849         & -36:53:18.57         & 2                    & CBD                  & \cite{Alencar+2003}\\
            \midrule
\multirow[m]{1}{*}{AS 205  $^2$              } & 16:11:31.346         & -18:38:25.96         & 3                    & CSD, CBD             & \cite{Ghez+1993,Eisner+2005,AndrewsWilliams2007}\\
            \midrule
\multirow[m]{1}{*}{BF Ori              } & 05:37:13.262         & -06:35:0.57          & 2                    & CSD                  & \cite{Stapper+2022,Thomas+2023}\\
            \midrule
\multirow[m]{1}{*}{BHB2007 11          } & 17:11:23.178         & -27:24:31.53         & 2                    & CSD,CBD              & \cite{Alves+2019}\\
            \midrule
\multirow[m]{1}{*}{Bernhard 2          } & 07:14:45.39          & -09:01:52.10         & 2                    & CBD                  & \cite{Zhu+2022,Hu+2024}\\
            \midrule
\multirow[m]{1}{*}{CD22 11432          } & 16:14:11.070         & -23:05:36.07         & 2                    & CSD/CBD              & \cite{MetchevHillenbrand2009,Kraus+2012,Barenfeld+2016,Davies2019}\\
            \midrule
\multirow[m]{1}{*}{CHX 22              } & 11:12:42.671         & -77:22:22.94         & 2                    & CBD                  & \cite{Daemgen+2013}\\
            \midrule
\multirow[m]{1}{*}{CIDA 9              } & 05:05:22.86          & +25:31:31.2          & 2                    & CSD                  & \cite{KrausHillenbrand2007,AkesonJensen2014}\\
            \midrule
\multirow[m]{1}{*}{CQ Tau              } & 05:35:58.467         & +24:44:54.09         & 2                    & CSD                  & \cite{Wheelwright+2011,Thomas+2023}\\
            \midrule
\multirow[m]{1}{*}{CS Cha              } & 11:02:24.876         & -77:33:35.67         & 2                    & CBD                  & \cite{Guenther+2007,Dunham+2016}\\
            \midrule
\multirow[m]{1}{*}{CoKu Tau/4          } & 04:41:16.810         & +28:40:00.0738       & 2                    & CBD                  & \cite{DAlessio+2005,IrelandKraus2008}\\
            \midrule
\multirow[m]{1}{*}{CoRoT 2239          } & 06:41:44.22          & +09:25:02.398        & 2                    & CBD                  & \cite{Gillen+2014}\\
            \midrule
\multirow[m]{1}{*}{DD Tau              } & 04:18:31.129         & +28:16:29.15         & 2                    & CSD                  & \cite{Bouvier+1992}\\
            \midrule
\multirow[m]{1}{*}{DF Tau              } & 04:27:2.793          & +25:42:22.45         & 2                    & CSD                  & \cite{WhiteGhez2001,Kutra+2024}\\
            \midrule
\multirow[m]{1}{*}{DH Tau              } & 04:29:41.659         & 26:32:56.508         & 2                    & CSD                  & \cite{Pascucci+2016}\\
            \midrule
\multirow[m]{1}{*}{DK Tau              } & 04:30:44.243         & +26:01:24.65         & 2                    & CSD                  & \cite{AkesonJensen2014}\\
            \midrule
\multirow[m]{1}{*}{DQ Tau              } & 04:46:53.058         & +17:00:00.14         & 2                    & CBD                  & \cite{Mathieu+1997,Guilloteau+2011}\\
            \midrule
\multirow[m]{1}{*}{DoAr 43             } & 16:31:30.882         & -24:24:39.888        & 2                    & CSD                  & \cite{Ratzka+2005,Cox+2017}\\
            \midrule
\multirow[m]{1}{*}{DoAr 51             } & 16:32:11.795         & -24:40:21.64         & 2                    & CSD                  & \cite{Ratzka+2005,Cox+2017}\\
            \midrule
\multirow[m]{1}{*}{EDJ2009-156         } & 03:28:51.029         & +31:18:18.409        & 2                    & CSD                  & \cite{Evans+2009,Tobin+2016}\\
            \midrule
\multirow[m]{1}{*}{EDJ2009-183         } & 03:28:59.297         & +31:15:48.410        & 2                    & CSD                  & \cite{Evans+2009,Tobin+2016}\\
            \midrule
\multirow[m]{1}{*}{EDJ2009-269         } & 03:30:44.014         & +30:32:46.812        & 2                    & CSD                  & \cite{Evans+2009,Tobin+2016}\\
            \midrule
\multirow[m]{1}{*}{FO Tau              } & 04:14:49.284         & +28:12:30.51         & 2                    & CSD                  & \cite{WhiteGhez2001,Akeson+2019}\\
            \midrule
\multirow[m]{1}{*}{FQ Tau              } & 04:19:12.813         & +28:29:32.98         & 2                    & CSD                  & \cite{Akeson+2019}\\
            \midrule
\multirow[m]{1}{*}{FU Ori  $^2$              } & 05:45:22.365         & +09:04:12.29         & 2                    & CSD                  & \cite{Wang+2004,Hales+2015}\\
            \midrule
\multirow[m]{1}{*}{FU Tau              } & 04:23:35.738         & 25:02:59.633         & 2                    & CSD                  & \cite{Pascucci+2016}\\
            \midrule
\multirow[m]{1}{*}{FV Tau              } & 04:26:53.537         & +26:06:54.28         & 4                    & CSD                  & \cite{AkesonJensen2014}\\
            \midrule
\multirow[m]{1}{*}{FX Tau              } & 04:30:29.618         & +24:26:45.03         & 2                    & CSD                  & \cite{AkesonJensen2014}\\
            \midrule
\multirow[m]{1}{*}{FZ Tau              } & 04:32:31.764         & 24:20:03.000         & 2                    & CSD                  & \cite{Pascucci+2016}\\
            \midrule
\multirow[m]{1}{*}{GG Tau              } & 04:32:30.351         & +17:31:40.49         & 5                    & CSD,CTD              & \cite{Leinert+1993,Dutrey+1994,DiFolco+2014,Roddier+1996,Silber+2000,Beck+2012,Keppler+2020}\\
            \midrule
\multirow[m]{1}{*}{GSS 31              } & 16:26:23.368         & -24:20:59.579        & 2                    & CSD                  & \cite{Ratzka+2005,Cox+2017}\\
            \midrule
\multirow[m]{1}{*}{GV Tau              } & 04:29:23.731         & +24:33:00.22         & 3                    & CSD,CBD              & \cite{Menard+1993,Doppmann+2008,Guilloteau+2011}\\
            \midrule
\multirow[m]{1}{*}{GW Ori              } & 05:29:08.393         & +11:52:12.67         & 3                    & CBD,CTD              & \cite{Mathieu+1991,Mathieu+1995,Berger+2011,Kraus+2020}\\
            \midrule
\multirow[m]{1}{*}{HBC 387             } & 04:26:54.401         & +26:06:50.95         & 2                    & CSD                  & \cite{AkesonJensen2014}\\
            \midrule
\multirow[m]{1}{*}{HBC 411             } & 04:35:40.938         & +24:11:8.49          & 2                    & CSD                  & \cite{AkesonJensen2014}\\
            \midrule
\multirow[m]{1}{*}{HBC 494             } & 05:40:27.450         & -07:27:30.06         & 2                    & CSD                  & \cite{Nogueira+2023}\\
            \midrule
\multirow[m]{1}{*}{HD 100453           } & 11:33:05.577         & -54:19:28.55         & 2                    & CSD                  & \cite{Chen+2006,Wagner+2015}\\
            \midrule
\multirow[m]{1}{*}{HD 104237           } & 12:00:05.087         & -78:11:34.57         & 2                    & CBD                  & \cite{Bohm+2004,Tatulli+2007}\\
            \midrule
\multirow[m]{1}{*}{HD 106906           } & 12:17:53.192         & -55:58:31.89         & 2                    & CBD                  & \cite{Kalas+2015,Lagrange+2016}\\
            \midrule
\multirow[m]{1}{*}{HD 131511           } & 14:53:23.765         & +19:09:10.07         & 2                    & CBD                  & \cite{Kennedy2015}\\
            \midrule
\multirow[m]{1}{*}{HD 141569           } & 15:49:57.748         & -03:55:16.34         & 3                    & CSD                  & \cite{Rich+2022,Thomas+2023}\\
            \midrule
\multirow[m]{1}{*}{HD 142527           } & 15:56:41.888         & -42:19:23.25         & 2                    & CSD,CBD              & \cite{Fukagawa+2006,Biller+2012,Marino+2015,Avenhaus+2017}\\
            \midrule
\multirow[m]{1}{*}{HD 144432           } & 16:06:57.953         & -27:43:9.76          & 3                    & CSD                  & \cite{Muller+2011,Varga+2024}\\
            \midrule
\multirow[m]{1}{*}{HD 150193           } & 16:40:17.924         & -23:53:45.19         & 2                    & CSD                  & \cite{Garufi+2020,Thomas+2023}\\
            \midrule
\multirow[m]{1}{*}{HD 200775           } & 21:01:36.921         & +68:09:47.79         & 3                    & CBD                  & \cite{Li+1994,Millan-Gabet+2001,Monnier+2006}\\
            \midrule
\multirow[m]{1}{*}{HD 34700            } & 05:19:41.409         & +05:38:42.78         & 3                    & CBD                  & \cite{Torres2004,Monnier+2019}\\
            \midrule
\multirow[m]{1}{*}{HD 36112            } & 05:30:27.529         & +25:19:57.08         & 2                    & CSD                  & \cite{Thomas+2023,Stapper+2024}\\
            \midrule
\multirow[m]{1}{*}{HD 37258            } & 05:36:59.250         & -06:09:16.33         & 2                    & CSD                  & \cite{Stapper+2022,Thomas+2023}\\
            \midrule
\multirow[m]{1}{*}{HD 98800            } & 11:22:05.290         & -24:46:39.76         & 4                    & CBD                  & \cite{Torres+1995,Koerner+2000}\\
            \midrule
\multirow[m]{1}{*}{HH 250              } & 19:21:13.675         & +10:52:31.00         & 2                    & CSD,CBD              & \cite{Comeron+2018}\\
            \midrule
\multirow[m]{1}{*}{HH 48               } & 11:04:22.800         & -77:18:07.99         & 2                    & CSD                  & \cite{Stapelfeldt+2014}\\
            \midrule
\multirow[m]{1}{*}{HK Ori              } & 05:31:28.050         & +12:09:10.30         & 3                    & CSD                  & \cite{Wheelwright+2011,Thomas+2023}\\
            \midrule
\multirow[m]{1}{*}{HK Tau              } & 04:31:50.572         & +24:24:17.78         & 2                    & CSD                  & \cite{Stapelfeldt+1998}\\
            \midrule
\multirow[m]{1}{*}{HN Tau              } & 04:33:39.363         & +17:51:52.29         & 2                    & CSD                  & \cite{Akeson+2019}\\
            \midrule
\multirow[m]{1}{*}{HP Cha              } & 11:08:15.216         & -77:33:53.17         & 3                    & CSD, CBD             & \cite{Daemgen+2013,Wolfer+2023}\\
            \midrule
\multirow[m]{1}{*}{HT Lup              } & 15:45:12.868         & -34:17:30.64         & 3                    & CSD                  & \cite{Ghez+1997,Andrews+2018}\\
            \midrule
\multirow[m]{1}{*}{HV Tau  $^2$              } & 04:38:35.290         & +26:10:38.64         & 3                    & CSD                  & \cite{Simon+1992,Stapelfeldt+2003}\\
            \midrule
\multirow[m]{1}{*}{Haro 6-37           } & 04:46:59.078         & 17:02:39.670         & 2                    & CSD                  & \cite{Pascucci+2016}\\
            \midrule
\multirow[m]{1}{*}{Hn 23               } & 13:04:24.060         & -76:50:01.390        & 2                    & CSD                  & \cite{Pascucci+2016}\\
            \midrule
\multirow[m]{1}{*}{IC348 LRL31         } & 03:44:18.17          & +32:04:57.00         & 2                    & CBD                  & \cite{Ruiz-Rodriguez+2016,Ruiz-Rodriguez+2018}\\
            \midrule
\multirow[m]{1}{*}{IC348 MMS           } & 03:43:57.065         & +32:03:04.788        & 3                    & CSD                  & \cite{Enoch+2009,Tobin+2016}\\
            \midrule
\multirow[m]{1}{*}{IRAS 03282+3035     } & 03:31:20.939         & +30:45:30.273        & 2                    & CSD                  & \cite{Enoch+2009,Tobin+2016}\\
            \midrule
\multirow[m]{1}{*}{IRAS 03282+30355    } & 03:44:43.298         & +32:01:31.236        & 3                    & CSD                  & \cite{Enoch+2009,Tobin+2016}\\
            \midrule
\multirow[m]{1}{*}{IRAS 03292+3039     } & 03:32:17.928         & +30:49:47.825        & 2                    & CSD                  & \cite{Enoch+2009,Tobin+2016}\\
            \midrule
\multirow[m]{1}{*}{IRAS 03292+30392    } & 03:25:22.409         & +30:45:13.258        & 2                    & CSD                  & \cite{Enoch+2009,Tobin+2016}\\
            \midrule
\multirow[m]{1}{*}{IRAS 04125+2902     } & 04:15:42.787         & +29:09:59.83         & 2                    & CSD                  & \cite{Barber+2024}\\
            \midrule
\multirow[m]{1}{*}{IRAS 04158+2805     } & 04:18:58.134         & +28:12:23.36         & 2                    & CSD,CBD              & \cite{Glauser+2008,Villenave+2020,Ragusa+2021}\\
            \midrule
\multirow[m]{1}{*}{IRAS 04298+2246     } & 04:32:49.117         & +22:53:2.85          & 4                    & CSD                  & \cite{AkesonJensen2014}\\
            \midrule
\multirow[m]{1}{*}{IRAS 16293-2422     } & 16:32:22.560         & -24:28:31.80         & 3                    & CSD,CBD              & \cite{Wootten1989,Maureira+2020}\\
            \midrule
\multirow[m]{1}{*}{IRAS 17216-3801     } & 17:25:6.517          & -38:04:0.44          & 2                    & CSD,CBD              & \cite{Kraus+2017}\\
            \midrule
\multirow[m]{1}{*}{IRAS F13052-7653N   } & 13:09:10.980         & -77:09:44.140        & 2                    & CSD                  & \cite{Pascucci+2016}\\
            \midrule
\multirow[m]{1}{*}{IRS 43              } & 16:27:26.950         & -24:40:50.500        & 2                    & CSD,CBD              & \cite{Girart+2000,Brinch+2016}\\
            \midrule
\multirow[m]{1}{*}{ISO-Oph 2   $^2$           } & 16:25:38.127         & -24:22:36.19         & 3                    & CSD                  & \cite{Cieza+2021}\\
            \midrule
\multirow[m]{1}{*}{IT Tau              } & 04:33:54.701         & +26:13:27.52         & 2                    & CSD                  & \cite{Akeson+2019}\\
            \midrule
\multirow[m]{1}{*}{J10555973-7724399   } & 10:55:59.731         & -77:24:39.913        & 2                    & CSD                  & \cite{Pascucci+2016}\\
            \midrule
\multirow[m]{1}{*}{J10574219-7659356   } & 10:57:42.200         & -76:59:35.664        & 2                    & CSD                  & \cite{Pascucci+2016}\\
            \midrule
\multirow[m]{1}{*}{J10581677-7717170   } & 10:58:16.774         & -77:17:17.056        & 2                    & CSD                  & \cite{Pascucci+2016}\\
            \midrule
\multirow[m]{1}{*}{J11023265-7729129   } & 11:02:32.654         & -77:29:12.977        & 2                    & CSD                  & \cite{Long+2018}\\
            \midrule
\multirow[m]{1}{*}{J11040909-7627193   } & 11:04:09.090         & -76:27:19.379        & 2                    & CSD                  & \cite{Pascucci+2016}\\
            \midrule
\multirow[m]{1}{*}{J11072074-7738073   } & 11:07:20.744         & -77:38:07.354        & 3                    & CSD                  & \cite{Pascucci+2016}\\
            \midrule
\multirow[m]{1}{*}{J11072825-7652118   } & 11:07:28.256         & -76:52:11.899        & 2                    & CSD                  & \cite{Long+2018}\\
            \midrule
\multirow[m]{1}{*}{J11075792-7738449   } & 11:07:57.928         & -77:38:44.927        & 2                    & CSD                  & \cite{Pascucci+2016}\\
            \midrule
\multirow[m]{1}{*}{J11080002-7717304   } & 11:08:00.025         & -77:17:30.484        & 2                    & CSD                  & \cite{Long+2018}\\
            \midrule
\multirow[m]{1}{*}{J11080148-7742288   } & 11:08:01.486         & -77:42:28.854        & 2                    & CSD                  & \cite{Pascucci+2016}\\
            \midrule
\multirow[m]{1}{*}{J11080297-7738425   } & 11:08:02.975         & -77:38:42.590        & 2                    & CSD                  & \cite{Pascucci+2016}\\
            \midrule
\multirow[m]{1}{*}{J11091812-7630292   } & 11:09:18.129         & -76:30:29.250        & 2                    & CSD                  & \cite{Long+2018}\\
            \midrule
\multirow[m]{1}{*}{J11095340-7634255   } & 11:09:53.405         & -76:34:25.511        & 2                    & CBD                  & \cite{Pascucci+2016}\\
            \midrule
\multirow[m]{1}{*}{J11095407-7629253   } & 11:09:54.076         & -76:29:25.310        & 2                    & CSD                  & \cite{Pascucci+2016}\\
            \midrule
\multirow[m]{1}{*}{J11095873-7737088   } & 11:09:58.738         & -77:37:08.879        & 2                    & CSD                  & \cite{Pascucci+2016}\\
            \midrule
\multirow[m]{1}{*}{J11100010-7634578   } & 11:10:00.108         & -76:34:57.893        & 2                    & CBD                  & \cite{Pascucci+2016}\\
            \midrule
\multirow[m]{1}{*}{J11100704-7629376   } & 11:10:07.045         & -76:29:37.698        & 2                    & CSD                  & \cite{Pascucci+2016}\\
            \midrule
\multirow[m]{1}{*}{J11103801-7732399   } & 11:10:38.018         & -77:32:39.905        & 2                    & CSD                  & \cite{Pascucci+2016}\\
            \midrule
\multirow[m]{1}{*}{J11105597-7645325   } & 11:10:55.974         & -76:45:32.573        & 2                    & CSD                  & \cite{Pascucci+2016}\\
            \midrule
\multirow[m]{1}{*}{J11122441-7637064   } & 11:12:24.415         & -76:37:06.406        & 2                    & CSD                  & \cite{Long+2018}\\
            \midrule
\multirow[m]{1}{*}{J11175211-7629392   } & 11:17:52.117         & -76:29:39.264        & 2                    & CSD                  & \cite{Pascucci+2016}\\
            \midrule
\multirow[m]{1}{*}{J15354856-2958551   } & 15:35:48.565         & -29:58:55.182        & 2                    & CSD                  & \cite{Barenfeld+2016}\\
            \midrule
\multirow[m]{1}{*}{J15534211-2049282   } & 15:53:42.119         & -20:49:28.218        & 3                    & CSD                  & \cite{Barenfeld+2016}\\
            \midrule
\multirow[m]{1}{*}{J16001844-2230114   } & 16:00:18.441         & -22:30:11.488        & 2                    & CSD                  & \cite{Barenfeld+2016}\\
            \midrule
\multirow[m]{1}{*}{J16014086-2258103   } & 16:01:40.869         & -22:58:10.384        & 2                    & CSD                  & \cite{Barenfeld+2016}\\
            \midrule
\multirow[m]{1}{*}{J16062196-1928445   } & 16:06:21.963         & -19:28:44.566        & 2                    & CSD                  & \cite{Carpenter+2014}\\
            \midrule
\multirow[m]{1}{*}{J16082751-1949047   } & 16:08:27.520         & -19:49:04.721        & 2                    & CSD                  & \cite{Carpenter+2014}\\
            \midrule
\multirow[m]{1}{*}{J16083070-3828268   } & 16:08:30.687         & -38:28:27.280        & 2                    & CSD                  & \cite{Ansdell+2016, Ansdell+2018}\\
            \midrule
\multirow[m]{1}{*}{J16084940-3905393   } & 16:08:49.382         & -39:05:39.825        & 2                    & CSD                  & \cite{Ansdell+2016, Ansdell+2018}\\
            \midrule
\multirow[m]{1}{*}{J16085373-3914367   } & 16:08:53.725         & -39:14:37.170        & 2                    & CSD                  & \cite{Pascucci+2016}\\
            \midrule
\multirow[m]{1}{*}{J16095628-3859518   } & 16:09:56.281         & -38:59:51.973        & 3                    & CSD                  & \cite{Ansdell+2016, Ansdell+2018}\\
            \midrule
\multirow[m]{1}{*}{J16101984-3836065   } & 16:10:19.840         & -38:36:06.800        & 2                    & CSD                  & \cite{Pascucci+2016}\\
            \midrule
\multirow[m]{1}{*}{J16133650-2503473   } & 16:13:36.510         & -25:03:47.340        & 2                    & CSD                  & \cite{Barenfeld+2016}\\
            \midrule
\multirow[m]{1}{*}{J16135434-2320342   } & 16:13:54.347         & -23:20:34.253        & 2                    & CSD                  & \cite{Barenfeld+2016}\\
            \midrule
\multirow[m]{1}{*}{KH15D  $^3$               } & 06:41:10.340         & +09:28:33.49         & 2                    & CBD                  & \cite{Hamilton+2001,Johnson+2004,ChiangMurray-Clay2004}\\
            \midrule
\multirow[m]{1}{*}{KK Oph              } & 17:10:8.110          & -27:15:19.01         & 2                    & CSD                  & \cite{Leinert+1997,Stapper+2022}\\
            \midrule
\multirow[m]{1}{*}{L1448 IRS1          } & 03:25:09.449         & +30:46:21.933        & 2                    & CSD                  & \cite{Tobin+2016}\\
            \midrule
\multirow[m]{1}{*}{L1448 IRS3          } & 03:25:36.379         & +30:45:14.728        & 6                    & CSD,CBD              & \cite{Tobin+2016}\\
            \midrule
\multirow[m]{1}{*}{L1455-FIR2          } & 03:27:38.268         & +30:13:58.448        & 2                    & CSD                  & \cite{Enoch+2009,Tobin+2016}\\
            \midrule
\multirow[m]{1}{*}{L1551 IRS5          } & 04:31:34.169         & +18:08:4.269         & 2                    & CBD                  & \cite{Reipurth2000,Reipurth+2002,Takakuwa+2012}\\
            \midrule
\multirow[m]{1}{*}{L1689 IRS5          } & 16:31:52.11          & -24:56:15.7          & 3                    & CSD                  & \cite{Ratzka+2005,Cox+2017}\\
            \midrule
\multirow[m]{1}{*}{MHO2                } & 04:14:26.401         & +28:05:59.64         & 3                    & CBD                  & \cite{Kraus+2011,Harris+2012}\\
            \midrule
\multirow[m]{1}{*}{NGC1333 IRAS1       } & 03:28:37.090         & +31:13:30.788        & 2                    & CSD                  & \cite{Enoch+2009,Tobin+2016}\\
            \midrule
\multirow[m]{1}{*}{NGC1333 IRAS2       } & 03:28:55.569         & +31:14:37.025        & 4                    & CSD                  & \cite{Rodriguez+1999,Enoch+2009,Tobin+2016}\\
            \midrule
\multirow[m]{1}{*}{NGC1333 IRAS4       } & 03:29:10.537         & +31:13:30.933        & 4                    & CSD                  & \cite{Enoch+2009,Tobin+2016}\\
            \midrule
\multirow[m]{1}{*}{NGC1333 IRAS7       } & 03:29:11.258         & +31:18:31.073        & 5                    & CSD                  & \cite{Enoch+2009,Tobin+2016}\\
            \midrule
\multirow[m]{1}{*}{Per-emb-17          } & 03:27:39.105         & +30:13:03.068        & 2                    & CSD                  & \cite{Enoch+2009,Tobin+2016}\\
            \midrule
\multirow[m]{1}{*}{Per-emb-40          } & 03:33:16.669         & +31:07:54.901        & 2                    & CSD                  & \cite{Enoch+2009,Tobin+2016}\\
            \midrule
\multirow[m]{1}{*}{ROXs 42C            } & 16:31:15.745         & -24:34:2.16          & 3                    & CBD                  & \cite{Mathieu+1989,Ghez+1993,Lee+1994,Barsony+2003}\\
            \midrule
\multirow[m]{1}{*}{ROph 36             } & 16:33:55.615         & -24:42:05.002        & 2                    & CBD                  & \cite{Ruiz-Rodriguez+2016,Cox+2017}\\
            \midrule
\multirow[m]{1}{*}{RW Aur  $^2$              } & 05:07:49.566         & +30:24:5.18          & 2                    & CSD                  & \cite{Cabrit+2006,Rodriguez+2018}\\
            \midrule
\multirow[m]{1}{*}{R CrA               } & 19:01:53.676         & -36:57:08.30         & 2                    & CBD                  & \cite{Kraus+2009,Mesa+2019}\\
            \midrule
\multirow[m]{1}{*}{SR 24  $^2$               } & 16:26:58.425         & -24:45:46.34         & 3                    & CSD,CBD              & \cite{Correia+2006,AndrewsWilliams2005,Mayama+2010}\\
            \midrule
\multirow[m]{1}{*}{SR 9                } & 16:27:40.29          & -24:22:04.0          & 2                    & CSD                  & \cite{Ratzka+2005,Cox+2017}\\
            \midrule
\multirow[m]{1}{*}{SVS 13              } & 03:29:03.764         & +31:16:03.808        & 5                    & CSD,CBD              & \cite{Enoch+2009,Tobin+2016}\\
            \midrule
\multirow[m]{1}{*}{SV Cep              } & 22:21:33.217         & +73:40:27.10         & 2                    & CSD                  & \cite{Stapper+2024,Thomas+2023}\\
            \midrule
\multirow[m]{1}{*}{S CrA               } & 19:01:08.597         & -36:57:19.90         & 2                    & CSD                  & \cite{Stapelfeldt+1997,Cazzoletti+2019}\\
            \midrule
\multirow[m]{1}{*}{Sz 123              } & 16:10:51.310         & -38:53:12.800        & 2                    & CSD                  & \cite{Ansdell+2016, Ansdell+2018}\\
            \midrule
\multirow[m]{1}{*}{Sz 65               } & 15:39:27.772         & -34:46:17.21         & 2                    & CSD                  & \cite{Miley+2024}\\
            \midrule
\multirow[m]{1}{*}{Sz 74               } & 15:48:05.213         & -35:15:53.342        & 2                    & CSD                  & \cite{Ansdell+2016, Ansdell+2018}\\
            \midrule
\multirow[m]{1}{*}{Sz 75               } & 15:49:12.086         & -35:39:05.463        & 2                    & CSD                  & \cite{Sanchis+2019, Ansdell+2018}\\
            \midrule
\multirow[m]{1}{*}{Sz 77               } & 15:51:46.941         & -35:56:44.531        & 2                    & CSD                  & \cite{Sanchis+2019, Ansdell+2018}\\
            \midrule
\multirow[m]{1}{*}{Sz 81               } & 15:55:50.264         & -38:01:34.087        & 2                    & CSD                  & \cite{Ansdell+2016, Ansdell+2018}\\
            \midrule
\multirow[m]{1}{*}{Sz 88               } & 16:07:00.582         & -39:02:19.913        & 3                    & CSD                  & \cite{Ansdell+2016, Ansdell+2018}\\
            \midrule
\multirow[m]{1}{*}{TMC 1               } & 04:41:45.900         & +25:41:26.99         & 2                    & CSD                  & \cite{vantHoff+2020}\\
            \midrule
\multirow[m]{1}{*}{TWA3A               } & 11:10:27.894         & -37:31:51.97         & 3                    & CBD                  & \cite{Muzerolle+2000,Torres+2003}\\
            \midrule
\multirow[m]{1}{*}{T Tau               } & 04:21:59.432         & +19:32:06.43         & 3                    & CSD,CBD              & \cite{WhiteGhez2001}\\
            \midrule
\multirow[m]{1}{*}{UX Tau  $^2$              } & 04:30:3.996          & +18:13:49.44         & 4                    & CSD                  & \cite{Duchene+1999,Correia+2006,Andrews+2011}\\
            \midrule
\multirow[m]{1}{*}{UY Aur              } & 04:51:47.389         & +30:47:13.55         & 2                    & CSD                  & \cite{Close+1998,Tang+2014}\\
            \midrule
\multirow[m]{1}{*}{UZ Tau              } & 04:32:43.022         & +25:52:30.90         & 4                    & CSD,CBD              & \cite{WhiteGhez2001,Jensen+1996}\\
            \midrule
\multirow[m]{1}{*}{V1685 Cyg           } & 20:20:28.241         & +41:21:51.53         & 2                    & CSD                  & \cite{Eisner+2004,Thomas+2023}\\
            \midrule
\multirow[m]{1}{*}{V347 Aur            } & 04:56:57.022         & +51:30:50.88         & 2                    & CBD                  & \cite{Donati+2024}\\
            \midrule
\multirow[m]{1}{*}{V4046 Sgr           } & 18:14:10.482         & -32:47:34.52         & 2                    & CBD                  & \cite{StempelsGahm2004,Rosenfeld+2012}\\
            \midrule
\multirow[m]{1}{*}{V710 Tau            } & 04:31:57.798         & +18:21:36.94         & 3                    & CSD                  & \cite{AkesonJensen2014}\\
            \midrule
\multirow[m]{1}{*}{V773 Tau  $^3$            } & 04:14:12.926         & +28:12:12.36         & 4                    & CSD                  & \cite{Ghez+1993,Duchene+2003,Torres+2012,Boden+2012}\\
            \midrule
\multirow[m]{1}{*}{V853 Oph            } & 16:28:45.277         & -24:28:18.91         & 3                    & CBD                  & \cite{Ratzka+2005,Cox+2017}\\
            \midrule
\multirow[m]{1}{*}{V856 Sco            } & 16:08:34.287         & -39:06:18.33         & 2                    & CSD                  & \cite{Ansdell+2018}\\
            \midrule
\multirow[m]{1}{*}{V892 Tau            } & 04:18:40.616         & +28:19:15.63         & 3                    & CBD                  & \cite{Smith+2005,Monnier+2008,Thomas+2023}\\
            \midrule
\multirow[m]{1}{*}{V935 Sco            } & 16:22:18.533         & -23:21:48.15         & 2                    & CBD                  & \cite{Ruiz-Rodriguez+2016,Cox+2017}\\
            \midrule
\multirow[m]{1}{*}{VLA 1623            } & 16:26:26.390         & -24:24:30.80         & 4                    & CSD,CBD              & \cite{Murillo+2013,Harris+2018}\\
            \midrule
\multirow[m]{1}{*}{WL4                 } & 16:27:18.487         & -24:29:05.91         & 3                    & CBD                  & \cite{Ratzka+2005,Plavchan+2008}\\
            \midrule
\multirow[m]{1}{*}{WL 20               } & 16:27:15.698         & -24:38:43.44         & 4                    & CSD,CBD              & \cite{Ressler+2000,Barsony+2024}\\
            \midrule
\multirow[m]{1}{*}{WSB 19              } & 16:25:02.2           & -24:59:31.0          & 2                    & CSD                  & \cite{Ratzka+2005,Cox+2017}\\
            \midrule
\multirow[m]{1}{*}{WSB 38              } & 16:26:46.429         & -24:12:00.079        & 3                    & CSD                  & \cite{Ratzka+2005,Cox+2017}\\
            \midrule
\multirow[m]{1}{*}{WSB 40              } & 16:26:48.659         & -23:56:34.14         & 2                    & CBD                  & \cite{Ruiz-Rodriguez+2016,Cox+2017}\\
            \midrule
\multirow[m]{1}{*}{WSB 74              } & 16:31:54.747         & -25:03:24.09         & 2                    & CBD                  & \cite{Kohn+2016,Cox+2017}\\
            \midrule
\multirow[m]{1}{*}{XY Per              } & 03:49:36.337         & +38:58:55.55         & 2                    & CSD                  & \cite{Stapper+2024}\\
            \midrule
\multirow[m]{1}{*}{YLW 16              } & 16:27:28.000         & -24:39:30.00         & 3                    & CBD                  & \cite{Herczeg+2011,Plavchan+2013}\\
            \midrule
\multirow[m]{1}{*}{Z CMa  $^2$               } & 07:03:43.160         & -11:33:06.21         & 2                    & CSD                  & \cite{Dong+2022}\\
            \midrule
\multirow[m]{1}{*}{alpha CrB           } & 15:34:41.268         & +26:42:52.87         & 2                    & CBD                  & \cite{Kennedy+2012}\\
            \midrule
\multirow[m]{1}{*}{beta Tri            } & 02:09:32.627         & +34:59:14.27         & 2                    & CBD                  & \cite{Kennedy+2012}\\
            \midrule
            \bottomrule

\multicolumn{6}{p{15cm}}{\footnotesize{The sample includes systems found in the literature with at least two confirmed stars separated by less than $1000$~au and with at least one disc that has been imaged. This table will be available in a machine-readable format. This table may not be exhaustive.  We encourage the community to share any additional systems or updated information to help complete this dataset. Contributions can be directed to the corresponding authors of this work.

$^1$ CSD : CircumStellar Disc, CBD : CircumBinary Disc, CTD : CircumTriple Disc. 

$^2$ System with an ongoing flyby event suspected, see \cite{Cuello+2023} for more details.

$^3$ Disc inferred via eclipse events.
}}
\end{longtable}
\end{adjustwidth}

%The appendix is an optional section that can contain details and data supplemental to the main text---for example, explanations of experimental details that would disrupt the flow of the main text but nonetheless remain crucial to understanding and reproducing the research shown; figures of replicates for experiments of which representative data are shown in the main text can be added here if brief, or as Supplementary Data. Mathematical proofs of results not central to the paper can be added as an appendix.

%%%%%%%%%%%%%%%%%%%%%%%%%%%%%%%%%%%%%%%%%%
\begin{adjustwidth}{-\extralength}{0cm}
\reftitle{References}


\bibliography{biblio.bib}
\end{adjustwidth}
\PublishersNote{}

%%%%%%%%%%%%%%%%%%%%%%%%%%%%%%%%%%%%%%%%%%

\end{document}